\documentclass[12pt,letterpaper]{article}
\usepackage{epsfig,rotating,setspace,latexsym,amsmath,epsf,amssymb,amsfonts,bm,theorem,cite,caption,subcaption,enumerate,longtable,accents}
\usepackage{algorithm,algorithmic,graphicx,epsf,authblk,epstopdf,url,color,multirow}
\usepackage{mathtools}
\usepackage{comment}
\usepackage{graphicx}

\newtheorem{theorem}{Theorem}

\newtheorem{definition}{Definition}
\newtheorem{remark}{Remark}
\newtheorem{lemma}{Lemma}

\allowdisplaybreaks

\begin{document}

\title{Deceptive Information Retrieval}

\author{Sajani Vithana \qquad Sennur Ulukus\\
\normalsize Department of Electrical and Computer Engineering\\
\normalsize University of Maryland, College Park, MD 20742 \\
\normalsize {\it spallego@umd.edu} \qquad {\it ulukus@umd.edu}}

\date{}
\maketitle

\vspace*{-1.0cm}

\begin{abstract}
We introduce the problem of deceptive information retrieval (DIR), in which a user wishes to download a required file out of multiple independent files stored in a system of databases while \emph{deceiving} the databases by making the databases' predictions on the user-required file index incorrect with high probability. Conceptually, DIR is an extension of private information retrieval (PIR). In PIR, a user downloads a required file without revealing its index to any of the databases. The metric of deception is defined as the probability of error of databases' prediction on the user-required file, minus the corresponding probability of error in PIR. The problem is defined on time-sensitive data that keeps updating from time to time. In the proposed scheme, the user deceives the databases by sending \emph{real} queries to download the required file at the time of the requirement and \emph{dummy} queries at multiple distinct future time instances to manipulate the probabilities of sending each query for each file requirement, using which the databases' make the predictions on the user-required file index. The proposed DIR scheme is based on a capacity achieving probabilistic PIR scheme, and achieves rates lower than the PIR capacity due to the additional downloads made to deceive the databases. When the required level of deception is zero, the proposed scheme achieves the PIR capacity. 
\end{abstract}

\section{Introduction}

Information is generally retrieved from a data storage system by directly requesting what is required. This is the most efficient form of information retrieval in terms of the download cost, as the user only downloads exactly what is required. However, if the user does not want to reveal the required information to the data storage system from which the information is retrieved, extra information must be requested to increase the uncertainty of the database's knowledge on the user's requirement. This is the core idea of private information retrieval (PIR) \cite{original,PIR,ChaoTian,coded,colluding,sideinfo,singleDB,byzantine,SecureStorage,XSTPIR,MMPIR,evesdroppers}, where the user downloads a required file out of $K$ independent files stored in $N$ non-colluding databases without revealing the required file index. In PIR, the databases' prediction of the user-required file based on the received queries is uniformly distributed across all files. Hence, the probability of error of the database's predictions in a PIR setting with $K$ files is $1-\frac{1}{K}$. In weakly private information retrieval \cite{leaky,ChaoTian_leakage}, a certain amount of information on the user-required file index is revealed to the databases to reduce the download cost. In such cases, as the databases have more information on the file index that the user requests, the error probability of the database's prediction is less than $1-\frac{1}{K}$. In this work, we study the case where the error probability of databases' prediction is larger than $1-\frac{1}{K}$. 

\begin{figure}
     \centering
         \includegraphics[scale=1]{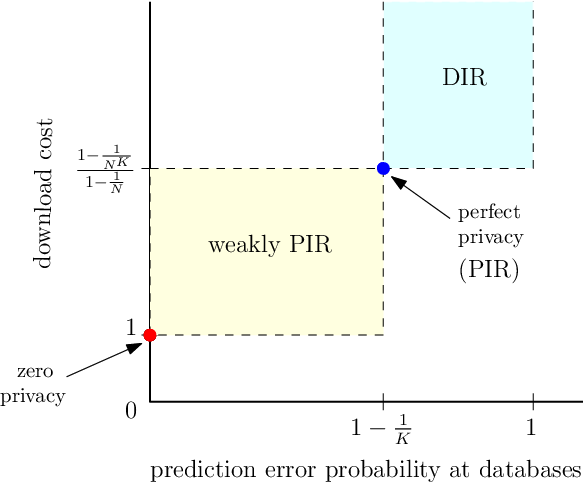}
          \caption{Download costs and prediction error probabilities for different types of information retrieval.}
          \label{fig:comp}
\end{figure}

Note that with no information received by the user at all, the databases can make a random guess on the user-required file index, and reach an error probability of $1-\frac{1}{K}$. Therefore, to result in a prediction error that is larger than $1-\frac{1}{K}$, the user has to \emph{deceive} the databases by sending fake information on the required file index. The goal of this work is to generate a scheme that allows a user to download a required file $k$, while forcing the databases' prediction on the user-required file index to be $\ell$, where $k\neq\ell$, for as many cases as possible. This is coined as deceptive information retrieval (DIR). DIR is achieved by sending \emph{dummy} queries to databases to manipulate the probabilities of sending each query for each file requirement, which results in incorrect predictions at the databases. However, sending dummy queries increases the download cost compared to PIR. Fig.~\ref{fig:comp} shows the behavior of the prediction error probability and the corresponding download costs for different types of information retrieval.\footnote{The regions marked as ``weakly PIR" and ``DIR" in Fig.~\ref{fig:comp} show the points that are conceptually valid for the two cases and does not imply that every point in those regions are achievable. The achievable points corresponding to ``weakly PIR" and ``DIR" lie within the marked regions.}

The concept of deception has been studied as a tool for cyber defense \cite{cyber_defense,cybersecurity,adaptivedeception,proofofconcept,diversity}, where the servers deceive attackers, adversaries and eavesdroppers to eliminate any harmful activities. In all such cases, the deceiver (servers in this case), gains nothing from the deceived, i.e., attackers, adversaries and eavesdroppers. In contrast, the main challenge in DIR is that what needs to be deceived is the same source of information that the user retrieves the required data from. This limits the freedom that a DIR scheme could employ to deceive the databases. To this end, we formulate the problem of DIR based on the key concepts used in PIR, while also incorporating a \emph{time dimension} to aid deception.

The problem of DIR introduced in this paper considers a system of non-colluding databases storing $K$ independent files that are time-sensitive, i.e., files that keep updating from time to time. We assume that the databases only store the latest version of the files. A given user wants to download arbitrary files at arbitrary time instances. The correctness condition ensures that the user receives the required file, right at the time of the requirement, while the condition for deception requires the databases' prediction on the user-required file to be incorrect with a probability that is greater than $1-\frac{1}{K}$, specified by the predetermined level of deception required in the system. 

The scheme that we propose for DIR deceives the databases by sending \emph{dummy} queries to the databases for each file requirement, at distinct time instances. From the user's perspective, each query is designed to play two roles as \emph{real} and \emph{dummy} queries, with two different probability distributions. This allows the user to manipulate the overall probability of sending each query for each message requirement, which is known by the databases. The databases make predictions based on the received queries and the globally known probability distribution of the queries used for each file requirement. These predictions are incorrect with probability $>1-\frac{1}{K}$ as the probability distributions based on which the real queries are sent are different from the globally known overall distribution. This is the basic idea used in the proposed scheme which allows a user to deceive the databases while also downloading the required file. 
The download cost of the proposed DIR scheme increases with the required level of deception $d$, and achieves the PIR capacity when $d=0$.

\section{Problem Formulation and System Model}\label{formulate}

We consider $N$ non-colluding databases storing $K$ independent files, each consisting of $L$ uniformly distributed symbols from a finite field $\mathbb{F}_q$, i.e.,
\begin{align}
    H(W_1,\dotsc,W_K)=\sum_{i=1}^K H(W_i)=KL,
\end{align}
where $W_i$ is the $i$th file. The files keep updating from time to time, and a given user wants to download an arbitrary file at arbitrary time instances $T_i$, $i\in\mathbb{N}$. We assume that all files are equally probable to be requested by the user.

The user sends queries at arbitrary time instances to download the required file while \emph{deceiving} the databases. We assume that the databases are only able to store data (files, queries from users, time stamps of received queries etc.) corresponding to the current time instance, and that the file updates at distinct time instances are mutually independent. Therefore, the user's file requirements and the queries sent are independent of the stored files at all time instances, i.e.,
\begin{align}
    I(\theta^{[t]},Q_n^{[t]};W_{1:K}^{[t]})=0, \quad n\in\{1,\dotsc,N\},\quad \forall t,
\end{align}
where $\theta^{[t]}$ is the user's file requirement, $Q_n^{[t]}$ is the query sent by the user to database $n$, and $W_{1:K}^{[t]}$ is the set of $K$ files, all at time $t$.\footnote{The notation $1:K$ indicates all integers from $1$ to $K$.} At any given time $t$ when each database $n$, $n\in\{1,\dotsc,N\}$, receives a query from the user, it sends the corresponding answer as a function of the received query and the stored files, thus,
\begin{align}
    H(A_n^{[t]}|Q_n^{[t]},W_{1:K}^{[t]})=0,\quad n\in\{1,\dotsc,N\},
\end{align}
where $A_n^{[t]}$ is the answer received by the user from database $n$ at time $t$. At each time $T_i$, $i\in\mathbb{N}$, the user must be able to correctly decode the required file, that is,
\begin{align}\label{correctness}
    H(W_{\theta^{[T_i]}}|Q_{1:N}^{[T_i]},A_{1:N}^{[T_i]})=0, \quad i\in\mathbb{N}.
\end{align}
At any given time $t$ when each database $n$, $n\in\{1,\dotsc,N\}$, receives a query from the user, it makes a prediction on the user-required file index using the maximum aposteriori probability (MAP) estimate as follows,
\begin{align}\label{predict}
    \hat{\theta}^{[t]}_{\Tilde{Q}}=\arg\max_{i} P(\theta^{[t]}=i|Q_n^{[t]}=\Tilde{Q}),\quad n\in\{1,\dotsc,N\},
\end{align}
where $\hat{\theta}^{[t]}_{\Tilde{Q}}$ is the predicted user-required file index based on the realization of the received query $\Tilde{Q}$ at time $t$. The probability of error of each database's prediction is defined as,
\begin{align}\label{perror}
    P_e=\mathbb{E}[P(\hat{\theta}^{[T_i]}_{\Tilde{Q}}\neq\theta^{[T_i]})],
\end{align}
where the expectation is taken across all $\Tilde{Q}$ and $T_i$. Note that in PIR, $P(\theta^{[t]}_{\Tilde{Q}}=i|Q_n^{[t]}=\Tilde{Q})=P(\theta^{[t]}_{\Tilde{Q}}=j|Q_n^{[t]}=\Tilde{Q})$ for all $i,j\in\{1,\dotsc,N\}$, any $\Tilde{Q}^{[t]}$, which results in $P_e^{\text{PIR}}=1-\frac{1}{K}$. Based on this information, we define the metric of deception as,
\begin{align}
    D=P_e-\left(1-\frac{1}{K}\right).
\end{align}
For PIR, the amount of deception is $D=0$, and for weakly PIR where some amount of information is leaked on the user-required file index, the amount of deception takes a negative value as the probability of error is smaller than $1-\frac{1}{K}$. The goal of this work is to generate schemes that meet a given level of deception $D=d>0$, while minimizing the normalized download cost defined as,
\begin{align}
    D_L=\frac{H(A_{1:N})}{L},
\end{align}
where $A_{1:N}$ represents all the answers received by all $N$ databases, corresponding to a single file requirement of the user. The DIR rate is defined as the reciprocal of $D_L$.

\section{Main Result}

In this section we present the main result of this paper, along with some remarks. Consider a system of $N$ non-colluding databases containing $K$ identical files. A user is able to retrieve any file $k$, while deceiving the databases by leaking information about some other file $k'$ to the databases.

\begin{theorem}
Consider a system of $N$ non-colluding databases storing $K$ independent files. A required level of deception $d$, satisfying $0\leq d<\frac{(K-1)(N-1)}{K(N^K-N)}$ is achievable at a DIR rate,
\begin{align}\label{main}
    R=\left(\frac{1+\left(\frac{N^K-N}{N-1}\right)e^\epsilon}{1+(N^{K-1}-1)e^\epsilon}+\left(\frac{N}{N-1}\right)(2u-u(u+1)\alpha)\right)^{-1},
\end{align}
where
\begin{align}
    \epsilon=\ln\left(\frac{dKN\!+\!(K\!-\!1)(N\!-\!1)}{dKN\!+\!(K\!-\!1)(N\!-\!1)\!-\!dKN^K}\right), \ \alpha=\frac{N\!+\!(N^K\!-\!N)e^{\epsilon}}{(N\!-\!1)e^{2\epsilon}\!+\!(N^K\!-\!N)e^{\epsilon}\!+\!1},\ u=\lfloor\frac{1}{\alpha}\rfloor 
\end{align}

\end{theorem}

\begin{remark}
For given $N$ and $K$, $\epsilon\geq0$ is a one-to-one continuous function of $d$, the required level of deception, and $\alpha\in(0,1]$ is a one-to-one continuous function of $\epsilon$. For a given $u\in\mathbb{Z^+}$, there exists a range of values of $\alpha$, specified by $\frac{1}{u+1}< \alpha\leq \frac{1}{u}$, which corresponds to a unique range of values of $\epsilon$, for which \eqref{main} is valid. Since $(0,1]=\cup\{{\alpha:\frac{1}{u+1}< \alpha\leq \frac{1}{u}, u\in\mathbb{Z^{+}}}\}$, there exists an achievable rate (as well as an achievable scheme) for any $\epsilon\geq0$ as well as for any $d$ in the range $0\leq d<\frac{(K-1)(N-1)}{K(N^K-N)}$.
\end{remark}

\begin{remark}
When the user specified amount of deception is zero, i.e., $d=0$, the corresponding values of $\alpha$ and $u$ are $\alpha=1$ and $u=1$. The achievable rate for this case is $\frac{1-\frac{1}{N}}{1-\frac{1}{N^K}}$, which is equal to the PIR capacity. 
\end{remark}

\begin{remark}
The achievable DIR rate monotonically decreases with increasing amount of deception $d$ for any given $N$ and $K$.
\end{remark}

\begin{remark}
    The variation of the achievable DIR rate with the level of deception for different number of databases when the number of files fixed at $K=3$ is shown in Fig.~\ref{varyingN}. The achievable rate for different number of files when the number of databases is fixed at $N=2$ is shown in Fig.~\ref{varyingK}. For any given $N$ and $K$, the rate decreases exponentially when the level of deception is close to the respective upper bound, i.e., $d<\frac{(K-1)(N-1)}{K(N^K-N)}$.

\begin{figure}
     \centering
         \includegraphics[scale=0.8]{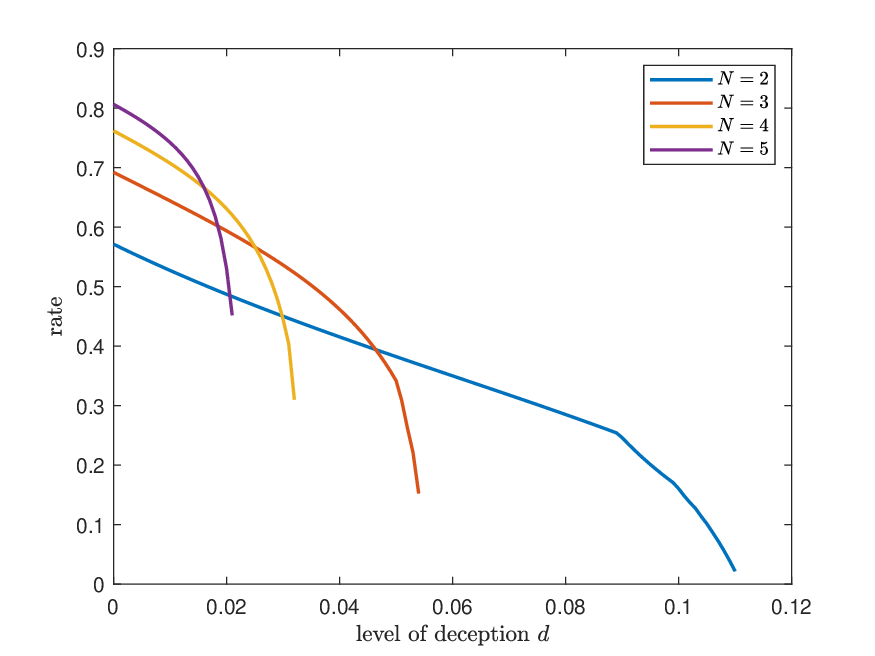}
          \caption{Achievable DIR rate for varying levels of deception and different number of databases when $K=3$.}
          \label{varyingN}
\end{figure}

\begin{figure}
     \centering
         \includegraphics[scale=0.8]{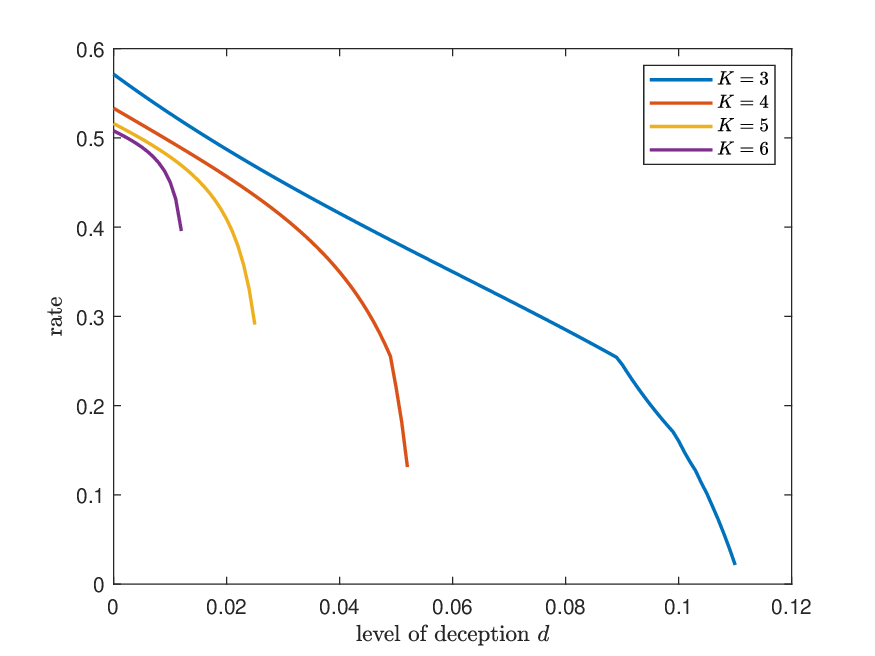}
          \caption{Achievable DIR rate for varying levels of deception and different number of files when $N=2$.}
          \label{varyingK}
\end{figure}
\end{remark}

\section{DIR Scheme}

The DIR scheme introduced in this section is designed for a system of $N$ non-colluding databases containing $K$ independent files, with a pre-determined amount of deception $d>0$ required. For each file requirement at time $T_i$, $i\in\mathbb{N}$, the user chooses a set of $M+1$ queries to be sent to database $n$, $n\in\{1,\dotsc,N\}$, at time $T_i$ as well as at future time instances $t_{i,j}$, $j\in\{1,\dotsc,M\}$, such that each $t_{i,j}>T_i$. The query sent at time $T_i$ is used to download the required file, while the rest of the $M$ queries are sent to deceive the databases. The queries sent at times $T_i$, $i\in\mathbb{N}$ and $t_{i,j}$, $j\in\{1,\dotsc,M\}$, $i\in\mathbb{N}$ are known as real and dummy queries, respectively. The binary random variable $R$ is used to specify whether a query sent by the user is real or dummy, i.e., $R=1$ corresponds to a real query sent at time $T_i$, and $R=0$ corresponds to a dummy query sent at time $t_{i,j}$. Next, we define another classification of queries used in the proposed scheme.

\begin{definition}[$\epsilon$-deceptive query]\label{def1}
An $\epsilon$-deceptive query $\Tilde{Q}$ with respect to file $k$ is defined as a query that always satisfies, 
\begin{align}\label{eq15}
    \frac{P(Q_n=\Tilde{Q}|\theta=k,R=1)}{P(Q_n=\Tilde{Q}|\theta=\ell,R=1)}=e^{-\epsilon}, \quad \frac{P(\theta=k|Q_n=\Tilde{Q})}{P(\theta=\ell|Q_n=\Tilde{Q})}=e^{\epsilon}, \quad \forall \ell\in \{1,\dotsc, K\}, \  \ell\neq k, 
\end{align}
for some $\epsilon>0$, where $Q_n$ and $\theta$ are the random variables representing a query sent to database $n$, $n\in\{1,\dotsc,N\}$, and the user-required file index. An equivalent representation of \eqref{eq15} is given by,
\begin{align}\label{eq16}
    \frac{P(R=1|\theta=\ell)+\frac{P(Q_n=\Tilde{Q}|\theta=\ell,R=0)}{P(Q_n=\Tilde{Q}|\theta=\ell,R=1)}P(R=0|\theta=\ell)}{P(R=1|\theta=k)+\frac{P(Q_n=\Tilde{Q}|\theta=k,R=0)}{P(Q_n=\Tilde{Q}|\theta=k,R=1)}P(R=0|\theta=k)}=e^{-2\epsilon}, \quad \forall \ell\in \{1,\dotsc, K\}, \  \ell\neq k.
\end{align}
\end{definition}

\begin{definition}[PIR query]\label{def2}
A query $\Tilde{Q}$ that satisfies \eqref{eq15} with $\epsilon=0$ for all $k\in\{1,\dotsc,K\}$, i.e., a $0$-deceptive query, is known as a PIR query.    
\end{definition}

\begin{remark}
    The intuition behind the definition of an $\epsilon$-deceptive query with respect to message $k$ in Definition~\ref{def1} is as follows. Note that the second equation in \eqref{eq15} fixes the databases’ prediction on the user’s requirement as $W_k$ for the query $\tilde{Q}$. This is because the aposteriori probability corresponding to message $k$, when $\tilde{Q}$ is received by the databases, is greater than that of any other message $\ell$, $\ell\neq k$. However, the first equation in \eqref{eq15}, which is satisfied at the same time, ensures that the user sends the query $\tilde{Q}$ with the least probability when the user requires to download message $k$, compared to the probabilities of sending $\tilde{Q}$ for other message requirements. In other words, since we assume equal priors, the query $\tilde{Q}$ is mostly sent when the user requires to download $W_\ell$ for $\ell\neq k$, and is rarely sent to download $W_k$, while the databases’ prediction on the user-required message upon receiving query $\tilde{Q}$ is fixed at $W_k$, which is incorrect with high probability, hence, the deception. 
\end{remark}

At a given time $t$, there exists a set of queries consisting of both deceptive and PIR queries, sent to the $N$ databases. Database $n$, $n\in\{1,\dotsc,N\}$, is aware of the probability of receiving each query, for each file requirement, i.e., $P(Q_n=\Tilde{Q}|\theta=k)$, for $k\in\{1,\dotsc,K\}$, $\Tilde{Q}\in\mathcal{Q}$, where $\mathcal{Q}$ is the set of all queries. However, the databases are unaware of being deceived, and are unable to determine whether the received query $\Tilde{Q}$ is real or dummy or deceptive or PIR. The proposed scheme generates a list of real and dummy queries for a given $N$ and $K$ along with the probabilities of using them as $\epsilon$-deceptive and PIR queries, based on the required level of deception $d$. The scheme also characterizes the optimum number of dummy queries $M$ to be sent to the databases for each file requirement, to minimize the download cost. As an illustration of the proposed scheme, consider the following representative examples.

\subsection{Example 1: Two Databases and Two Files, $N=K=2$}

In this example, we present how the proposed DIR scheme is applied in a system of two databases containing two files each. In the proposed scheme, the user generates $M+1$ queries for any given file-requirement which consists of one real query and $M$ dummy queries. The user sends the real query at the time of the requirement $T_i$, and the rest of the $M$ dummy queries at $M$ different future time instances $t_{i,j}$. Tables~\ref{tab1} and~\ref{tab2} give possible pairs of real queries that are sent to the two databases to retrieve $W_1$ and $W_2$, respectively, at time $T_i$, $i\in\mathbb{N}$. The probability of using each pair of queries is indicated in the first columns of Tables~\ref{tab1} and ~\ref{tab2}. Note that the correctness condition in \eqref{correctness} is satisfied at each time $T_i$ as each row of Tables~\ref{tab1} and~\ref{tab2} decodes files $W_1$ and $W_2$, respectively, with no error. 

\begin{table}[ht]
\begin{minipage}{0.5\textwidth}
\begin{center}
\begin{tabular}{ |c|c|c|}
\hline
 $P(Q|\theta=1,R=1)$ & DB 1 & DB 2\\
\hline
$p$ & $W_1$ & $\phi$ \\ 
\hline
$p$ & $\phi$ & $W_1$ \\ 
\hline
$p'$ & $W_2$ & $W_1+W_2$ \\
\hline
 $p'$ & $W_1+W_2$ & $W_2$ \\
\hline
\end{tabular}
\end{center}
\vspace*{-0.4cm}
\caption{Real query table -- $W_1$.}
\label{tab1}
\end{minipage}
\hfill
\begin{minipage}{0.5\textwidth}
\begin{center}
\begin{tabular}{ |c|c|c|}
\hline
 $P(Q|\theta=2,R=1)$ & DB 1 & DB 2\\
\hline
$p$ & $W_2$ & $\phi$ \\ 
\hline
$p$ & $\phi$ & $W_2$ \\ 
\hline
$p'$ & $W_1$ & $W_1+W_2$ \\
\hline
 $p'$ & $W_1+W_2$ & $W_1$ \\
\hline
\end{tabular}
\end{center}
\vspace*{-0.4cm}
\caption{Real query table -- $W_2$.}
\label{tab2}
\end{minipage}
\end{table}

\begin{table}[ht]
\begin{minipage}{0.5\textwidth}
\begin{center}
\begin{tabular}{ |c|c|c|}
\hline
 $P(Q|\theta=1,R=0)$ & DB 1 & DB 2\\
\hline
$1$ & $W_1$ & $W_1$ \\ 
\hline
\end{tabular}
\end{center}
\vspace*{-0.4cm}
\caption{Dummy query table -- $W_1$.}
\label{tab3}
\end{minipage}
\hfill
\begin{minipage}{0.5\textwidth}
\begin{center}
\begin{tabular}{ |c|c|c|}
\hline
 $P(Q|\theta=2,R=0)$ & DB 1 & DB 2\\
\hline
$1$ & $W_2$ & $W_2$ \\ 
\hline
\end{tabular}
\end{center}
\vspace*{-0.4cm}
\caption{Dummy query table -- $W_2$.}
\label{tab4}
\end{minipage}
\end{table}

The dummy queries sent to each database at time $t_{i,j}$ are given in Tables~\ref{tab3} and~\ref{tab4}.  The purpose of the dummy queries sent at future time instances is to deceive the databases by manipulating the aposteriori probabilities, which impact their predictions. For example, if the user wants to download $W_1$ at time $T_i$, the user selects one of the four query options in Table~\ref{tab1} based on the probabilities in column 1,\footnote{The values of $p$ and $p'$ are derived later in this section.} and sends the corresponding queries to database 1 and 2 at time $T_i$. Based on the information in Table~\ref{tab3}, the user sends the query $W_1$ to both databases at $M$ distinct future time instances $t_{i,j}$, $j\in\{1,\dotsc,M\}$.

Based on the information in Tables~\ref{tab1}-\ref{tab4}, when the user-required file is $W_1$, the probability of each query being received by database $n$, $n\in\{1,2\}$, at an arbitrary time instance $t$ is calculated as follows. Let $P(R=1|\theta=i)=\alpha$ for $i\in\{1,2\}$.\footnote{The intuition behind $P(R=1|\theta=i)$ is the probability of a query received by any database being real when the user-required file index is $i$. For a fixed $M$, $P(R=1|\theta=i)=\frac{1}{M+1}$.} Then, 
\begin{align}
    P(Q_n=W_1|\theta=1)&=P(Q_n=W_1|\theta=1,R=1)P(R=1|\theta=1)\nonumber\\
    &\quad+P(Q_n=W_1|\theta=1,R=0)P(R=0|\theta=1)\\
    &=p\alpha+1-\alpha\\
    P(Q_n=W_2|\theta=1)&=P(Q_n=W_2|\theta=1,R=1)P(R=1|\theta=1)\nonumber\\
    &\quad+P(Q_n=W_2|\theta=1,R=0)P(R=0|\theta=1)\\
    &=p'\alpha\\
    P(Q_n=W_1+W_2|\theta=1)&=P(Q_n=W_1+W_2|\theta=1,R=1)P(R=1|\theta=1)\nonumber\\
    &\quad+P(Q_n=W_1+W_2|\theta=1,R=0)P(R=0|\theta=1)\\
    &=p'\alpha\\
     P(Q_n=\phi|\theta=1)&=P(Q_n=\phi|\theta=1,R=1)P(R=1|\theta=1)\nonumber\\
    &\quad+P(Q_n=\phi|\theta=1,R=0)P(R=0|\theta=1)\\
    &=p\alpha
\end{align}
Thus, writing these probabilities compactly, we have,
\begin{align}\label{first1}
    P(Q_n=W_1|\theta=1)&=p\alpha+1-\alpha\\
    P(Q_n=W_2|\theta=1)&=p'\alpha\\
    P(Q_n=W_1+W_2|\theta=1)&=p'\alpha\\
     P(Q_n=\phi|\theta=1)&=p\alpha.
\end{align}
Similarly, when the user-required file is $W_2$, the corresponding probabilities are,
\begin{align}
    P(Q_n=W_1|\theta=2)&=p'\alpha\\
    P(Q_n=W_2|\theta=2)&=p\alpha+1-\alpha\\
    P(Q_n=W_1+W_2|\theta=2)&=p'\alpha\\
     P(Q_n=\phi|\theta=2)&=p\alpha.\label{last1}
\end{align}

These queries and the corresponding probabilities of sending them to each database for each message requirement are known to the databases. However, the decomposition of these probabilities based on whether the query is real or dummy, i.e., Tables~\ref{tab1}-\ref{tab4}, is not known by the databases. When database $n$, $n\in\{1,\dotsc,N\}$, receives a query $\Tilde{Q}$ at time $t$, it calculates the aposteriori probability distribution of the user-required file index, to predict the user's requirement using \eqref{predict}. The aposteriori probabilities corresponding to the four queries received by database $n$, $n\in\{1,2\}$, are calculated as follows,
\begin{align}\label{posterior}
    P(\theta=i|Q_n=\Tilde{Q})&=\frac{P(Q_n=\Tilde{Q}|\theta=i)P(\theta=i)}{P(Q_n=\Tilde{Q})}.
\end{align}
Then, the explicit a posteriori probabilities are given by,
\begin{align}\label{pos1}
    P(\theta=1|Q_n=W_1)&=\frac{\frac{1}{2}(p\alpha+1-\alpha)}{P(Q_n=W_1)}\\
    P(\theta=2|Q_n=W_1)&=\frac{\frac{1}{2}p'\alpha}{P(Q_n=W_1)}\\
    P(\theta=1|Q_n=W_2)&=\frac{\frac{1}{2}p'\alpha}{P(Q_n=W_2)}\\
    P(\theta=2|Q_n=W_2)&=\frac{\frac{1}{2}(p\alpha+1-\alpha)}{P(Q_n=W_2)}\\
    P(\theta=1|Q_n=W_1+W_2)&=\frac{\frac{1}{2}p'\alpha}{P(Q_n=W_1+W_2)}\\
    P(\theta=2|Q_n=W_1+W_2)&=\frac{\frac{1}{2}p'\alpha}{P(Q_n=W_1+W_2)}\\
    P(\theta=1|Q_n=\phi)&=\frac{\frac{1}{2}p\alpha}{P(Q_n=\phi)}\\
    P(\theta=2|Q_n=\phi)&=\frac{\frac{1}{2}p\alpha}{P(Q_n=\phi)}.\label{poslast}
\end{align}

While queries $\phi$ and $W_1+W_2$ are PIR queries as stated in Definition~\ref{def2}, queries $W_1$ and $W_2$ are $\epsilon$-deceptive with respect to file indices $1$ and $2$, respectively, for an $\epsilon$ that depends on the required amount of deception $d$. The values of $p$ and $p'$ in Tables~\ref{tab1}-\ref{tab4} are calculated based on the requirements in Definition~\ref{def1} as follows. It is straightforward to see that $p'=pe^\epsilon$ follows from the first part of \eqref{eq15} for each query $\Tilde{Q}=W_1$ and $\Tilde{Q}=W_2$, which also gives $p=\frac{1}{2(1+e^{\epsilon})}$. The second part of \eqref{eq15} (as well as \eqref{eq16}) results in $\alpha=\frac{2}{1+e^{\epsilon}}$ for both $\epsilon$-deceptive queries $W_1$ and $W_2$. Based on the aposteriori probabilities \eqref{pos1}-\eqref{poslast} calculated by the databases using the information in \eqref{first1}-\eqref{last1}, each database predicts the user's requirement at each time it receives a query from the user. The predictions corresponding to each query received by database $n$, $n=1,2$, which are computed using \eqref{predict}, are shown in Table~\ref{dbpredicts}. 

\begin{table}[ht]
\begin{center}
\begin{tabular}{ |c|c|c|}
\hline
query $\Tilde{Q}$ & $P(\hat{\theta}_{\Tilde{Q}}=1)$ & $P(\hat{\theta}_{\Tilde{Q}}=2)$\\
\hline
$W_1$ & $1$ & $0$\\ 
\hline
$W_2$ & $0$ & $1$\\
\hline
$W_1+W_2$ & $\frac{1}{2}$ & $\frac{1}{2}$\\
\hline
$\phi$ & $\frac{1}{2}$ & $\frac{1}{2}$\\
\hline
\end{tabular}
\end{center}
\caption{Probabilities of each database predicting the user-required file in Example~1.}
\label{dbpredicts}
\end{table}

Based on this information, when a database receives query $Q=W_1$, it always decides that the requested message is $W_1$, and when it receives query $Q=W_2$, it always decides that the requested message is $W_2$. For queries $Q=\phi$ and $Q=W_1+W_2$, the databases flip a coin to choose either $W_1$ or $W_2$ as the requested message. 

As the queries are symmetric across all databases, the probability of error corresponding to some query $\Tilde{Q}$ received by database $n$ at time $T_i$ is given by,
\begin{align}  
   &P(\hat{\theta}^{[T_i]}_{\Tilde{Q}}\neq\theta^{[T_i]}) \nonumber\\
   &=P(\theta^{[T_i]}=1,\hat{\theta}_{\Tilde{Q}}^{[T_i]}= 2|Q_n^{[T_i]}=\Tilde{Q})+P(\theta^{[T_i]}=2,\hat{\theta}^{[T_i]}_{\Tilde{Q}}=1|Q_n^{[T_i]}=\Tilde{Q})\\
   &=\frac{1}{P(Q_n^{[T_i]}=\Tilde{Q})}\left(P(\hat{\theta}^{[T_i]}_{\Tilde{Q}}=2|\theta^{[T_i]}=1,Q_n^{[T_i]}=\Tilde{Q})P(Q_n^{[T_i]}=\Tilde{Q}|\theta^{[T_i]}=1)P(\theta^{[T_i]}=1)\right.\nonumber\\
    &\qquad \qquad \qquad \quad \left.+ P(\hat{\theta}_{\Tilde{Q}}^{[T_i]}=1|\theta^{[T_i]}=2,Q_n^{[T_i]}=\Tilde{Q})P(Q_n^{[T_i]}=\Tilde{Q}|\theta^{[T_i]}=2)P(\theta^{[T_i]}=2)\right) \\
    &=\frac{1}{P(Q_n^{[T_i]}=\Tilde{Q})}\left(P(\hat{\theta}^{[T_i]}_{\Tilde{Q}}=2|Q_n^{[T_i]}=\Tilde{Q})P(Q_n^{[T_i]}=\Tilde{Q}|\theta^{[T_i]}=1)P(\theta^{[T_i]}=1)\right.\nonumber\\
    &\qquad \qquad \quad \qquad\left.+P(\hat{\theta}_{\Tilde{Q}}=1|Q_n^{[T_i]}=\Tilde{Q})P(Q_n^{[T_i]}=\Tilde{Q}|\theta^{[T_i]}=2)P(\theta^{[T_i]}=2)\right),
\end{align}
as the predictions only depend on the received queries. The explicit probabilities corresponding to the four queries are,\footnote{Note that $P(Q_n=\Tilde{Q}|\theta^{[T_i]}=i)$ implies $P(Q_n=\Tilde{Q}|\theta=i,R=1)$ as only real queries are sent at time $T_i$.}
\begin{align}
    P(\hat{\theta}_{W_1}^{[T_i]}\neq\theta^{[T_i]})&=\frac{1}{P(Q_n^{[T_i]}=W_1)}\frac{e^\epsilon}{4(1+e^\epsilon)}\\
    P(\hat{\theta}_{W_2}^{[T_i]}\neq\theta^{[T_i]})&=\frac{1}{P(Q_n^{[T_i]}=W_2)}\frac{e^\epsilon}{4(1+e^\epsilon)}\\
    P(\hat{\theta}_{W_1+W_2}^{[T_i]}\neq\theta^{[T_i]})&=\frac{1}{P(Q_n^{[T_i]}=W_1+W_2)}\frac{e^\epsilon}{4(1+e^\epsilon)}\\
    P(\hat{\theta}_{\phi}^{[T_i]}\neq\theta^{[T_i]})&=\frac{1}{P(Q_n^{[T_i]}=\phi)}\frac{1}{4(1+e^\epsilon)}.
\end{align}
As the same scheme is used for all user-requirements at all time instances, the probability of error of each database's prediction for this example is calculated using \eqref{perror} as,
\begin{align}
    P_e&=\sum_{\Tilde{Q}\in\mathcal{Q}} P(Q_n^{[T_i]}=\Tilde{Q})P(\hat{\theta}_{\Tilde{Q}}^{[T_i]}\neq\theta^{[T_i]}) \\
    &=\frac{3e^\epsilon+1}{4(1+e^\epsilon)}
\end{align}
where $\mathcal{Q}=\{W_1,W_2,W_1+W_2,\phi\}$, which results in a deception of $D=\frac{3e^\epsilon+1}{4(1+e^\epsilon)}-\frac{1}{2}=\frac{e^\epsilon-1}{4(1+e^\epsilon)}$. Therefore, for a required amount of deception $d<\frac{1}{4}$, the value of $\epsilon$ is chosen as $\epsilon=\ln\left(\frac{4d+1}{1-4d}\right)$. 

The download cost of this scheme is computed as follows. As the scheme is symmetric across all file retrievals, and since the apriori probability distribution of the files is uniform, without loss of generality, we can calculate the download cost of retrieving $W_1$. The download cost of retrieving $W_1$ for a user specified amount of deception $d$ is given by,
\begin{align}
    D_L&=\frac{1}{L}\left(2Lp+2(2L)pe^\epsilon+2L\sum_{m=0}^{\infty} p_mm\right)\\
    &=\frac{1+2e^\epsilon}{1+e^\epsilon}+2\mathbb{E}[M]
\end{align}
where $p_m$ is the probability of sending $m$ dummy queries per each file requirement. To minimize the download cost, we need to find the probability mass function (PMF) of $M$ which minimizes $\mathbb{E}[M]$ such that $P(R=1|\theta=i)=\alpha=\frac{2}{1+e^\epsilon}$ is satisfied for any $i$. Note that for any 
$i$, $P(R=1|\theta=i)$ can be written as,
\begin{align}\label{expect}
    P(R=1|\theta=i)=\alpha=\sum_{m=0}^\infty p_m\frac{1}{m+1}=\mathbb{E}\left[\frac{1}{M+1}\right], 
\end{align}
where $M$ is the random variable representing the number of dummy queries sent to each database per file requirement. Thus, the following optimization problem needs to be solved, for a given $\epsilon$, that is a function of the given value of $d$,
\begin{align}
    \min & \quad \mathbb{E}[M] \nonumber\\
    \text{s.t.} & \quad  \mathbb{E}\left[\frac{1}{M+1}\right]=\frac{2}{1+e^{\epsilon}}=\alpha.
\end{align}
The solution to this problem is given in Lemma~\ref{Lemma1}, and the resulting minimum download cost is given by,
\begin{align}
    D_L&=\frac{1+2e^\epsilon}{1+e^\epsilon}+4u-2u(u+1)\alpha,
\end{align}
where $u=\lfloor\frac{1}{\alpha}\rfloor$. When $d=0$, it follows that $\epsilon=0$ and $u=1$, and the achievable rate is $\frac{2}{3}$, which is the same as the PIR capacity for $N=2$ and $K=2$. 

\subsection{Example 2: Three Databases and Three Files, $N=K=3$}

Similar to the previous example, the user sends real queries at time $T_i$ and dummy queries at times $t_{i,j}$, $j\in\{1,\dotsc,M\}$, for each $i\in\mathbb{N}$, based on the probabilities shown in Tables~\ref{real1}-\ref{dummy3}. The notation $W_i^j$ in these tables correspond to the $j$th segment of $W_i$, where each file $W_i$ is divided into $N-1=2$ segments of equal size. Database $n$, $n\in\{1,\dotsc,N\}$, only knows the overall probabilities of receiving each query for each file requirement of the user shown in Table~\ref{dbknows2}. These overall probabilities which are calculated using, 
\begin{align}\label{scheme}    P(Q_n=\Tilde{Q}|\theta=k)&=P(Q_n=\Tilde{Q}|\theta=k,R=1)P(R=1|\theta=k)\nonumber\\
    &\quad+P(Q_n=\Tilde{Q}|\theta=k,R=0)P(R=0|\theta=k), \quad k\in\{1,\dotsc,K\}
\end{align}
where $P(R=1|\theta=i)=\alpha$ for any $i=1,2,3$, are the same for each database as the scheme is symmetric across all databases. 

\begin{table}[ht]
\begin{center}
\begin{tabular}{ |c|c|c|c| }
\hline
query $\Tilde{Q}$ & $P(Q_n=\Tilde{Q}|\theta=1)$ &  $P(Q_n=\Tilde{Q}|\theta=2)$ & $P(Q_n=\Tilde{Q}|\theta=3)$  \\
\hline
$\phi$ & $p\alpha$ & $p\alpha$ & $p\alpha$\\
$W_1^1$ & $p\alpha+\frac{1}{2}(1-\alpha)$ & $p'\alpha$ & $p'\alpha$\\
$W_1^2$ & $p\alpha+\frac{1}{2}(1-\alpha)$ & $p'\alpha$ & $p'\alpha$\\
$W_2^1$ & $p'\alpha$ & $p\alpha+\frac{1}{2}(1-\alpha)$ & $p'\alpha$\\
$W_2^2$ & $p'\alpha$ & $p\alpha+\frac{1}{2}(1-\alpha)$ & $p'\alpha$\\
$W_3^1$ & $p'\alpha$ & $p'\alpha$ & $p\alpha+\frac{1}{2}(1-\alpha)$\\
$W_3^2$ & $p'\alpha$ & $p'\alpha$ & $p\alpha+\frac{1}{2}(1-\alpha)$\\
other queries & $p'\alpha$  & $p'\alpha$ & $p'\alpha$\\
\hline
\end{tabular}
\end{center}
\vspace*{-0.4cm}
\caption{Queries received by database $n$, $n\in\{1,\dotsc,N\}$, at a given time $t$ for each file requirement, and the corresponding probabilities.}
\label{dbknows2}
\end{table}
The entry ``other queries" in Table~\ref{dbknows2} includes all queries that have sums of two or three elements. Based on this available information, each database calculates the aposteriori probability of the user-required file index conditioned on each received query $\Tilde{Q}$ using \eqref{posterior}. Each query of the form $W_k^j$ is an $\epsilon$-deceptive query with respect to file $k$, where $\epsilon$ is a function of the required amount of deception, which is derived towards the end of this section. All other queries including the null query and all sums of two or three elements are PIR queries. As all $\epsilon$-deceptive queries must satisfy \eqref{eq15}, the value of $p'$ is given by $p'=pe^\epsilon$, which results in $p=\frac{1}{3(1+8e^\epsilon)}$, based on the same arguments used in the previous example. Using \eqref{eq15} and \eqref{posterior} for any given deceptive query, the value of $\alpha$ is calculated as follows. Note that for a query of the form $W_k^j$, for each database $n$, $n\in\{1,\dotsc,N\}$, using $P(\theta=k)=\frac{1}{K}$, we have
\begin{align}\label{calcpost2}
    \frac{P(\theta=k|Q_n=W_k^j)}{P(\theta=\ell|Q=W_k^j)}&=\frac{P(Q_n=W_k^j|\theta=k)}{P(Q_n=W_k^j|\theta=\ell)}=\frac{p\alpha+\frac{1}{2}(1-\alpha)}{p'\alpha},
\end{align}
The value of $\alpha$ is computed as $\alpha=\frac{1}{2p(e^{2\epsilon}-1)+1}$, using \eqref{calcpost2} and \eqref{eq15} by solving $\frac{p\alpha+\frac{1}{2}(1-\alpha)}{p'\alpha}=e^\epsilon$. 

\begin{table}[ht]
\begin{center}
\begin{tabular}{ |c|c|c|c|c| }
\hline
$P(Q|\theta=1,R=1)$ &  Database 1 & Database 2 & Database 3 \\
\hline
$p$ & $W_1^1$ & $W_1^2$ & $\phi$\\
$p$ & $W_1^2$ & $\phi$ & $W_1^1$\\
$p$ & $\phi$ & $W_1^1$ & $W_1^2$\\
\hline
$p'$ & $W_1^1+W_2^1$ & $W_1^2+W_2^1$ & $W_2^1$\\
$p'$ & $W_1^2+W_2^1$ & $W_2^1$ & $W_1^1+W_2^1$\\
$p'$ & $W_2^1$ & $W_1^1+W_2^1$ & $W_1^2+W_2^1$\\
\hline
$p'$ & $W_1^1+W_2^2$ & $W_1^2+W_2^2$ & $W_2^2$\\
$p'$ & $W_1^2+W_2^2$ & $W_2^2$ & $W_1^1+W_2^2$\\
$p'$ & $W_2^2$ & $W_1^1+W_2^2$ & $W_1^2+W_2^2$\\
\hline
$p'$ & $W_1^1+W_3^1$ & $W_1^2+W_3^1$ & $W_3^1$\\
$p'$ & $W_1^2+W_3^1$ & $W_3^1$ & $W_1^1+W_3^1$\\
$p'$ & $W_3^1$ & $W_1^1+W_3^1$ & $W_1^2+W_3^1$\\
\hline
$p'$ & $W_1^1+W_3^2$ & $W_1^2+W_3^2$ & $W_3^2$\\
$p'$ & $W_1^2+W_3^2$ & $W_3^2$ & $W_1^1+W_3^2$\\
$p'$ & $W_3^2$ & $W_1^1+W_3^2$ & $W_1^2+W_3^2$\\
\hline
$p'$ & $W_1^1+W_2^1+W_3^1$ & $W_1^2+W_2^1+W_3^1$ & $W_2^1+W_3^1$\\
$p'$ & $W_1^2+W_2^1+W_3^1$ & $W_2^1+W_3^1$ & $W_1^1+W_2^1+W_3^1$\\
$p'$ & $W_2^1+W_3^1$ & $W_1^1+W_2^1+W_3^1$ & $W_1^2+W_2^1+W_3^1$\\
\hline
$p'$ & $W_1^1+W_2^2+W_3^1$ & $W_1^2+W_2^2+W_3^1$ & $W_2^2+W_3^1$\\
$p'$ & $W_1^2+W_2^2+W_3^1$ & $W_2^2+W_3^1$ & $W_1^1+W_2^2+W_3^1$\\
$p'$ & $W_2^2+W_3^1$ & $W_1^1+W_2^2+W_3^1$ & $W_1^2+W_2^2+W_3^1$\\
\hline
$p'$ & $W_1^1+W_2^1+W_3^2$ & $W_1^2+W_2^1+W_3^2$ & $W_2^1+W_3^2$\\
$p'$ & $W_1^2+W_2^1+W_3^2$ & $W_2^1+W_3^2$ & $W_1^1+W_2^1+W_3^2$\\
$p'$ & $W_2^1+W_3^2$ & $W_1^1+W_2^1+W_3^2$ & $W_1^2+W_2^1+W_3^2$\\
\hline
$p'$ & $W_1^1+W_2^2+W_3^2$ & $W_1^2+W_2^2+W_3^2$ & $W_2^2+W_3^2$\\
$p'$ & $W_1^2+W_2^2+W_3^2$ & $W_2^2+W_3^2$ & $W_1^1+W_2^2+W_3^2$\\
$p'$ & $W_2^2+W_3^2$ & $W_1^1+W_2^2+W_3^2$ & $W_1^2+W_2^2+W_3^2$\\
\hline
\end{tabular}
\end{center}
\vspace*{-0.4cm}
\caption{Real query table -- $W_1$.}
\label{real1}
\end{table}

\begin{table}[h!]
\begin{center}
\begin{tabular}{ |c|c|c|c|c|c|}
\hline
$P(Q|\theta=1,R=0)$ & DB 1 & $P(Q|\theta=1,R=0)$ & DB 2 & $P(Q|\theta=1,R=0)$ & DB 3\\
\hline
$\frac{1}{2}$ & $W_1^1$ & $\frac{1}{2}$ & $W_1^1$ & $\frac{1}{2}$ & $W_1^1$ \\ 
\hline
$\frac{1}{2}$ & $W_1^2$ & $\frac{1}{2}$ & $W_1^2$ & $\frac{1}{2}$ & $W_1^2$\\
\hline
\end{tabular}
\end{center}
\vspace*{-0.4cm}
\caption{Dummy query table -- $W_1$.}
\label{dummy1}
\end{table}

\begin{table}[ht]
\begin{center}
\begin{tabular}{ |c|c|c|c|c| }
\hline
 $P(Q|\theta=2,R=1)$ &  Database 1 & Database 2 & Database 3 \\
\hline
$p$ & $W_2^1$ & $W_2^2$ & $\phi$\\
$p$ & $W_2^2$ & $\phi$ & $W_2^1$\\
$p$ & $\phi$ & $W_2^1$ & $W_2^2$\\
\hline
$p'$ & $W_1^1+W_2^1$ & $W_1^1+W_2^2$ & $W_1^1$\\
$p'$ & $W_1^1+W_2^2$ & $W_1^1$ & $W_1^1+W_2^1$\\
$p'$ & $W_1^1$ & $W_1^1+W_2^1$ & $W_1^1+W_2^2$\\
\hline
$p'$ & $W_1^2+W_2^1$ & $W_1^2+W_2^2$ & $W_1^2$\\
$p'$ & $W_1^2+W_2^2$ & $W_1^2$ & $W_1^2+W_2^1$\\
$p'$ & $W_1^2$ & $W_1^2+W_2^1$ & $W_1^2+W_2^2$\\
\hline
$p'$ & $W_2^1+W_3^1$ & $W_2^2+W_3^1$ & $W_3^1$\\
$p'$ & $W_2^2+W_3^1$ & $W_3^1$ & $W_2^1+W_3^1$\\
$p'$ & $W_3^1$ & $W_2^1+W_3^1$ & $W_2^2+W_3^1$\\
\hline
$p'$ & $W_2^1+W_3^2$ & $W_2^2+W_3^2$ & $W_3^2$\\
$p'$ & $W_2^2+W_3^2$ & $W_3^2$ & $W_2^1+W_3^2$\\
$p'$ & $W_3^2$ & $W_2^1+W_3^2$ & $W_2^2+W_3^2$\\
\hline
$p'$ & $W_1^1+W_2^1+W_3^1$ & $W_1^1+W_2^2+W_3^1$ & $W_1^1+W_3^1$\\
$p'$ & $W_1^1+W_2^2+W_3^1$ & $W_1^1+W_3^1$ & $W_1^1+W_2^1+W_3^1$\\
$p'$ & $W_1^1+W_3^1$ & $W_1^1+W_2^1+W_3^1$ & $W_1^1+W_2^2+W_3^1$\\
\hline
$p'$ & $W_1^1+W_2^1+W_3^2$ & $W_1^1+W_2^2+W_3^2$ & $W_1^1+W_3^2$\\
$p'$ & $W_1^1+W_2^2+W_3^2$ & $W_1^1+W_3^2$ & $W_1^1+W_2^1+W_3^2$\\
$p'$ & $W_1^1+W_3^2$ & $W_1^1+W_2^1+W_3^2$ & $W_1^1+W_2^2+W_3^2$\\
\hline
$p'$ & $W_1^2+W_2^1+W_3^1$ & $W_1^2+W_2^2+W_3^1$ & $W_1^2+W_3^1$\\
$p'$ & $W_1^2+W_2^2+W_3^1$ & $W_1^2+W_3^1$ & $W_1^2+W_2^1+W_3^1$\\
$p'$ & $W_1^2+W_3^1$ & $W_1^2+W_2^1+W_3^1$ & $W_1^2+W_2^2+W_3^1$\\
\hline
$p'$ & $W_1^2+W_2^1+W_3^2$ & $W_1^2+W_2^2+W_3^2$ & $W_1^2+W_3^2$\\
$p'$ & $W_1^2+W_2^2+W_3^2$ & $W_1^2+W_3^2$ & $W_1^2+W_2^1+W_3^2$\\
$p'$ & $W_1^2+W_3^2$ & $W_1^2+W_2^1+W_3^2$ & $W_1^2+W_2^2+W_3^2$\\
\hline
\end{tabular}
\end{center}
\vspace*{-0.4cm}
\caption{Real query table -- $W_2$.}
\label{real2}
\end{table}

\begin{table}[h!]
\begin{center}
\begin{tabular}{ |c|c|c|c|c|c|}
\hline
$P(Q|\theta=2,R=0)$ & DB 1 & $P(Q|\theta=2,R=0)$ & DB 2 & $P(Q|\theta=2,R=0)$ & DB 3\\
\hline
$\frac{1}{2}$ & $W_2^1$ & $\frac{1}{2}$ & $W_2^1$ & $\frac{1}{2}$ & $W_2^1$ \\ 
\hline
$\frac{1}{2}$ & $W_2^2$ & $\frac{1}{2}$ & $W_2^2$ & $\frac{1}{2}$ & $W_2^2$\\
\hline
\end{tabular}
\end{center}
\vspace*{-0.4cm}
\caption{Dummy query table -- $W_2$.}
\label{dummy2}
\end{table}

\begin{table}[ht]
\begin{center}
\begin{tabular}{ |c|c|c|c|c| }
\hline
$P(Q|\theta=3,R=1)$ &  Database 1 & Database 2 & Database 3 \\
\hline
$p$ & $W_3^1$ & $W_3^2$ & $\phi$\\
$p$ & $W_3^2$ & $\phi$ & $W_3^1$\\
$p$ & $\phi$ & $W_3^1$ & $W_3^2$\\
\hline
$p'$ & $W_1^1+W_3^1$ & $W_1^1+W_3^2$ & $W_1^1$\\
$p'$ & $W_1^1+W_3^2$ & $W_1^1$ & $W_1^1+W_3^1$\\
$p'$ & $W_1^1$ & $W_1^1+W_3^1$ & $W_1^1+W_3^2$\\
\hline
$p'$ & $W_1^2+W_3^1$ & $W_1^2+W_3^2$ & $W_1^2$\\
$p'$ & $W_1^2+W_3^2$ & $W_1^2$ & $W_1^2+W_3^2$\\
$p'$ & $W_1^2$ & $W_1^2+W_3^2$ & $W_1^2+W_3^1$\\
\hline
$p'$ & $W_2^1+W_3^1$ & $W_2^1+W_3^2$ & $W_2^1$\\
$p'$ & $W_2^1+W_3^2$ & $W_3^1$ & $W_2^1+W_3^1$\\
$p'$ & $W_2^1$ & $W_2^1+W_3^1$ & $W_2^1+W_3^2$\\
\hline
$p'$ & $W_2^2+W_3^1$ & $W_2^2+W_3^2$ & $W_2^2$\\
$p'$ & $W_2^2+W_3^2$ & $W_2^2$ & $W_2^2+W_3^1$\\
$p'$ & $W_2^2$ & $W_2^2+W_3^1$ & $W_2^2+W_3^2$\\
\hline
$p'$ & $W_1^1+W_2^1+W_3^1$ & $W_1^1+W_2^1+W_3^2$ & $W_1^1+W_2^1$\\
$p'$ & $W_1^1+W_2^1+W_3^2$ & $W_1^1+W_2^1$ & $W_1^1+W_2^1+W_3^1$\\
$p'$ & $W_1^1+W_2^1$ & $W_1^1+W_2^1+W_3^1$ & $W_1^1+W_2^1+W_3^2$\\
\hline
$p'$ & $W_1^2+W_2^1+W_3^1$ & $W_1^2+W_2^1+W_3^2$ & $W_1^2+W_2^1$\\
$p'$ & $W_1^2+W_2^1+W_3^2$ & $W_1^2+W_2^1$ & $W_1^2+W_2^1+W_3^1$\\
$p'$ & $W_1^2+W_2^1$ & $W_1^2+W_2^1+W_3^1$ & $W_1^2+W_2^1+W_3^2$\\
\hline
$p'$ & $W_1^1+W_2^2+W_3^1$ & $W_1^1+W_2^2+W_3^2$ & $W_1^1+W_2^2$\\
$p'$ & $W_1^1+W_2^2+W_3^2$ & $W_1^1+W_2^2$ & $W_1^1+W_2^2+W_3^1$\\
$p'$ & $W_1^1+W_2^2$ & $W_1^1+W_2^2+W_3^1$ & $W_1^1+W_2^2+W_3^2$\\
\hline
$p'$ & $W_1^2+W_2^2+W_3^1$ & $W_1^2+W_2^2+W_3^2$ & $W_1^2+W_2^2$\\
$p'$ & $W_1^2+W_2^2+W_3^2$ & $W_1^2+W_2^2$ & $W_1^2+W_2^2+W_3^1$\\
$p'$ & $W_1^2+W_2^2$ & $W_1^2+W_2^2+W_3^1$ & $W_1^2+W_2^2+W_3^2$\\
\hline
\end{tabular}
\end{center}
\vspace*{-0.4cm}
\caption{Real query table -- $W_3$.}
\label{real3}
\end{table}

\begin{table}[h!]
\begin{center}
\begin{tabular}{ |c|c|c|c|c|c|}
\hline
$P(Q|\theta=3,R=0)$ & DB 1 & $P(Q|\theta=3,R=0)$ & DB 2 & $P(Q|\theta=3,R=0)$ & DB 3\\
\hline
$\frac{1}{2}$ & $W_3^1$ & $\frac{1}{2}$ & $W_3^1$ & $\frac{1}{2}$ & $W_3^1$ \\ 
\hline
$\frac{1}{2}$ & $W_3^2$ & $\frac{1}{2}$ & $W_3^2$ & $\frac{1}{2}$ & $W_3^2$\\
\hline
\end{tabular}
\end{center}
\vspace*{-0.4cm}
\caption{Dummy query table -- $W_3$.}
\label{dummy3}
\end{table}

Assume that the user wants to download $W_2$ at some time $T_i$. Then, at time $T_i$, the user picks a row of queries from Table~\ref{real2} based on the probabilities in the first column, and sends them to each of the three databases. Note that correctness is satisfied as it is possible to decode $W_2$ from any row of Table~\ref{real2}. Next, the user picks $M$ future time instances $t_{i,j}$, $j\in\{1,\dotsc,M\}$, and at each time $t_{i,j}$ the user independently and randomly picks a row from Table~\ref{dummy2} and sends the queries to the databases. This completes the scheme, and the value of $M$ that minimizes the download cost is calculated at the end of this example. 

The databases make predictions with the received query at each time $t$, based on the information available in Table~\ref{dbknows2}. As the aposteriori probabilities $P(\theta=k|Q_n=\Tilde{Q})$ are proportional to the corresponding probabilities given by $P(Q_n=\Tilde{Q}|\theta=k)$ from \eqref{posterior}, the databases' predictions (using \eqref{predict}) and the corresponding probabilities are shown in Table~\ref{dbpredicts2}.

\begin{table}[ht]
\begin{center}
\begin{tabular}{ |c|c|c|c|}
\hline
query $\Tilde{Q}$ & $P(\hat{\theta}_{\Tilde{Q}}=1)$ & $P(\hat{\theta}_{\Tilde{Q}}=2)$ & $P(\hat{\theta}_{\Tilde{Q}}=3)$ \\
\hline
$W_1^1$ & $1$ & $0$ & $0$\\ 
\hline
$W_1^2$ & $1$ & $0$ & $0$\\
\hline
$W_2^1$ & $0$ & $1$ & $0$\\
\hline
$W_2^2$ & $0$ & $1$ & $0$\\
\hline
$W_3^1$ & $0$ & $0$ & $1$\\
\hline
$W_3^2$ & $0$ & $0$ & $1$\\
\hline
other queries & $\frac{1}{3}$ & $\frac{1}{3}$ & $\frac{1}{3}$\\
\hline
\end{tabular}
\end{center}
\caption{Probabilities of each database predicting the user-required file in Example~2.}
\label{dbpredicts2}
\end{table}

The probability of error for each type of query is calculated as follows. First, consider the $\epsilon$-deceptive queries with respect to file $k$, given by $W_k^j$, $j\in\{1,2\}$. For these queries, the error probability from the perspective of database $n$, $n\in\{1,\dotsc,N\}$, is given by,
\begin{align}
    P(\hat{\theta}^{[T_i]}_{W_k^j}\neq\theta^{[T_i]})&=P(\theta^{[T_i]}\neq k|Q_n^{[T_i]}=W_k^j)\label{expla1}\\
    &=\sum_{\ell=1,\ell\neq k}^3 P(\theta^{[T_i]}= \ell|Q_n^{[T_i]}=W_k^j)\\
    &=\frac{\sum_{\ell=1,\ell\neq k}^3 P(Q_n^{[T_i]}=W_k^j|\theta^{[T_i]}= \ell)P(\theta^{[T_i]}= \ell)}{P(Q_n^{[T_i]}=W_k^j)}\\
    &=\frac{1}{P(Q_n^{[T_i]}=W_k^j)}\frac{2}{3}pe^\epsilon,\label{expla2}
\end{align}
where \eqref{expla1} follows from the fact that the databases' prediction on a received query of the form $W_k^j$ is file $k$ with probability 1 from Table~\ref{dbpredicts2}, and the probabilities in \eqref{expla2} are obtained from real query tables as they correspond to queries sent at time $T_i$. Next, the probability of error corresponding to each of the the other queries, i.e., PIR queries that include the null query and sums of two or three elements, is given by,
\begin{align}
    P(\hat{\theta}^{[T_i]}_{\Tilde{Q}}\neq\theta^{[T_i]})&\!=\!P(\hat{\theta}^{[T_i]}\neq\theta^{[T_i]}|Q_n^{[T_i]}\!=\!\Tilde{Q})\\    
    &\!=\!\frac{\sum_{j=1}^3 \sum_{m=1,m\neq j}^3 P(\hat{\theta}^{[T_i]}\!=\!m,\theta^{[T_i]}\!=\!j,Q_n^{[T_i]}\!=\!\Tilde{Q})}{P(Q_n^{[T_i]}\!=\!\Tilde{Q})}\\
    &\!=\!\frac{\sum_{j=1}^3 \sum_{m=1,m\neq j}^3 P(\hat{\theta}^{[T_i]}\!=\!m|\theta^{[T_i]}\!=\!j,Q_n^{[T_i]}\!=\!\Tilde{Q})P(Q_n^{[T_i]}\!=\!\Tilde{Q}|\theta^{[T_i]}\!=\!j)P(\theta^{[T_i]}\!=\!j)}{P(Q_n^{[T_i]}\!=\!\Tilde{Q})}\\
    &\!=\!\frac{1}{P(Q_n^{[T_i]}\!=\!\Tilde{Q})}\begin{cases}
        \frac{2p}{3}, & \text{if } \Tilde{Q}=\phi\\
        \frac{2pe^\epsilon}{3}, & \text{if $\Tilde{Q}$ if of the form $\sum_{s=1}^\ell W_{k_s}^{j_s}$ for $\ell\in\{2,3\}$} 
    \end{cases}\label{expla3}
\end{align}
where \eqref{expla3} follows from the fact that $\hat{\theta}^{[T_i]}$ is conditionally independent of $\theta^{[T_i]}$ given $Q_n$, from \eqref{predict}. 
The probability of error at each time $T_i$, $i\in\mathbb{N}$, is the same, as the scheme is identical at each $T_i$, and across all file requirements. Therefore, the probability of error of each database's prediction, using \eqref{perror} is given by,
\begin{align}
    P_e&=P(\hat{\theta}^{[T_i]}\neq\theta^{[T_i]})\\
    &=\sum_{\Tilde{Q}\in\mathcal{Q}}P(Q_n=\Tilde{Q})P(\hat{\theta}^{[T_i]}_{\Tilde{Q}}\neq\theta^{[T_i]})\\
    &=\sum_{k=1}^3\sum_{j=1}^{2}P(Q_n=W_k^j)\frac{1}{P(Q_n^{[T_i]}=W_k^j)}\frac{2}{3}pe^\epsilon+P(Q_n=\phi)\frac{1}{P(Q_n=\phi)}\frac{2p}{3}\nonumber\\
    &\qquad+20P(Q_n=\hat{Q})\frac{1}{P(Q_n=\hat{Q})}\frac{2pe^\epsilon}{3}\\ 
    &=4pe^\epsilon+\frac{2p}{3}+\frac{40pe^\epsilon}{3}\\
    &=\frac{52e^\epsilon+2}{9(8e^\epsilon+1)}.
\end{align}
where $\mathcal{Q}$ is the set of all queries and $\hat{Q}$ is a query of the form $\sum_{s=1}^\ell W_{k_s}^{j_s}$ for $\ell\in\{2,3\}$. The resulting amount of deception is,
\begin{align}
    D&=P_e-\left(1-\frac{1}{K}\right)=\frac{52e^\epsilon+2}{9(8e^\epsilon+1)}-\frac{2}{3}=\frac{4(e^\epsilon-1)}{9(8e^\epsilon+1)}.
\end{align}
Therefore, for a required amount of deception $d<\frac{1}{18}$, $\epsilon$ is chosen as $\epsilon=\ln\left(\frac{9d+4}{4(1-18d)}\right)$. 

Without loss of generality, consider the cost of downloading $W_1$, which is the same as the expected download cost, as the scheme is symmetric across all file retrievals,
\begin{align}
    D_L&=\frac{1}{L}\left(L\times3p+\frac{3L}{2}\times24pe^\epsilon+\frac{3L}{2}\sum_{m=0}^\infty p_m m\right)=\frac{1+12e^\epsilon}{1+8e^\epsilon}+\frac{3}{2}\mathbb{E}[M]
\end{align}
To find the scheme that achieves the minimum $D_L$ we need to find the minimum $\mathbb{E}[M]$ that satisfies $P(R=1|\theta=i)=\alpha=\mathbb{E}[\frac{1}{M+1}]=\frac{3(1+8e^\epsilon)}{2e^{2\epsilon}+24e^\epsilon+1}$, i.e., the following optimization problem needs to be solved.
\begin{align}
    \min & \quad \mathbb{E}[M]\nonumber\\
    \text{s.t.} & \quad  \mathbb{E}\left[\frac{1}{M+1}\right]=\frac{3e^{-2\epsilon}(1+8e^\epsilon)}{2+e^{-2\epsilon}+24e^{-\epsilon}}.
\end{align}
The solution to this problem is given in Lemma~\ref{Lemma1}. The resulting minimum download cost for a given value of $\epsilon$, i.e., required level of deception $d$, is given by,
\begin{align}
     \frac{D_\epsilon}{L}&=\frac{1+12e^\epsilon}{1+8e^\epsilon}+\frac{3}{2}(2u-u(u+1)\alpha), \quad \alpha=\frac{3e^{-2\epsilon}(1+8e^\epsilon)}{2+e^{-2\epsilon}+24e^{-\epsilon}},
\end{align}
where $u=\lfloor\frac{1}{\alpha}\rfloor$. When $d=0$, it follows that $\epsilon=0$, $\alpha=1$ and $u=1$, and the achievable rate is $\frac{9}{13}$, which is equal the PIR capacity for the case $N=3,K=3$.  

\subsection{Generalized DIR Scheme for Arbitrary $N$ and $K$ }

In the general DIR scheme proposed in this work, at each time $T_i$, $i\in\mathbb{N}$, when the user requires to download some file $W_k$, the user sends a set of real queries to each of the $N$ databases. These queries are picked based on a certain probability distribution, defined on all possible sets of real queries. For the same file requirement, the user sends $M$ dummy queries at future time instances $t_{i,j}$, $j\in\{1,\dotsc,M\}$, where $t_{i,j}>T_i$. The dummy queries sent at each time $t_{i,j}$ are randomly selected from a subset of real queries. We assume that the databases are unaware of being deceived, and treat both real and dummy queries the same when calculating their predictions on the user-required file index at each time they receive a query. The overall probabilities of a given user sending each query for each file requirement is known by the databases. However, the decomposition of these probabilities based on whether each query is used as a real or a dummy query is not known by the databases. It is also assumed that the databases only store the queries received at the current time instance.

The main components of the general scheme include 1) $N^K$ possible sets of real queries to be sent to the $N$ databases for each file requirement and their probabilities, 2) $N-1$ possible sets of dummy queries and their probabilities, 3) overall probabilities of sending each query for each of the $K$ file requirements of the user. Note that 1) and 2) are only known by the user while 3) is known by the databases. 

As shown in the examples considered, the set of all possible real queries takes the form of the queries in the probabilistic PIR scheme in \cite{ravi-leaky,semanticPIR}, with a non-uniform probability distribution unlike in PIR. The real query table used when retrieving $W_k$ consists of the following queries:

\begin{enumerate}
\item\textbf{Single blocks:} $W_k$ is divided into $N-1$ parts, and each part is requested from $N-1$ databases, while requesting nothing $\phi$ from the remaining database. All cyclic shifts of these queries are considered in the real query table.

\item\textbf{Sums of two blocks/Single block:} One database is used to download $W_j^l$, $l\in\{1,\dotsc,N-1\},j\neq k$ and each one in the rest of the $N-1$ databases is used to download $W_k^r+W_j^l$ for each $r\in\{1,\dotsc,N-1\}$. All cyclic shifts of these queries are also considered as separate possible sets of queries.

\item\textbf{Sums of three/Two blocks:} One database is used to download $W_{j_1}^{\ell_1}+W_{j_2}^{\ell_2}$, $ \ell_1,\ell_2\in\{1,\dotsc,N-1\}$ and $j_1\neq j_2\neq k$. Each one in the rest of the $N-1$ databases is used to download $W_{j_1}^{l_1}+W_{j_2}^{l_2}+W_k^r$ for each $r\in\{1,\dotsc,N-1\}$. All cyclic shifts of these queries are also considered as separate possible sets of queries.

\item\textbf{Sums of $K$/$K-1$ blocks:} The above process is repeated for all sums of blocks until $K$/$K-1$.
\end{enumerate}
Out of the $N^K$ different sets of queries described above in the real query table, the queries except $\phi$ in single blocks, i.e., queries of the form $W_k^\ell$, $\ell\in\{1,\dotsc,N-1\}$, are chosen as $\epsilon$-deceptive ones with respect to file $k$, for each $k\in\{1,\dotsc,K\}$, and are included in the set of dummy queries sent to databases when the user-required file index is $k$. The $N-1$ $\epsilon$-deceptive queries $W_k^r$, $r\in\{1,\dotsc,N-1\}$, corresponding to the $k$th file requirement must guarantee the condition in \eqref{eq15}. For that, we assign,
\begin{align}
    P(Q_n=W_k^r|\theta=k,R=1)=p, \quad r\in\{1,\dotsc,N-1\}
\end{align}
and
\begin{align}
    P(Q_n=W_k^r|\theta=j,R=1)=pe^\epsilon, \quad r\in\{1,\dotsc,N-1\},\quad j\neq k,
\end{align}
for each database $n$, $n\in\{1,\dotsc,N\}$. The rest of the queries, i.e., $\phi$ and sums of $\ell$ blocks where $\ell\in\{2,\dotsc,K\}$, are PIR queries in the proposed scheme. Note that the query $\phi$ is always coupled with the $\epsilon$-deceptive queries with respect to file index $k$ (required file) for correctness (see Tables~\ref{real1},~\ref{real2},~\ref{real3}). Thus, $\phi$ is assigned the corresponding probability given by,
\begin{align}
    P(Q_n=\phi|\theta=m,R=1)=p, \quad m\in\{1,\dotsc,K\}, \quad n\in\{1,\dotsc,N\}.    
\end{align}
Similarly, as the rest of the PIR queries are coupled with $\epsilon$-deceptive queries with respect to file indices $j$, $j\neq k$, or with other PIR queries, they are assigned the corresponding probability given by,
\begin{align}\label{pirquery}
    P(Q_n=\hat{Q}|\theta=m,R=1)=pe^\epsilon, \quad m\in\{1,\dotsc,K\}, \quad n\in\{1,\dotsc,N\},    
\end{align}
where $\hat{Q}$ is any PIR query in the form of $\ell$-sums with $\ell\in\{2,\dotsc,K\}$. Since the probabilities of the real queries sent for each file requirement must add up to one, i.e., $\sum_{\Tilde{Q}\in\mathcal{Q}} P(Q_n=\Tilde{Q}|\theta=m,R=1)=1$ for each $m\in\{1,\dotsc,K\}$, $p$ is given by,
\begin{align}
    p=\frac{1}{N+(N^K-N)e^\epsilon},
\end{align}
as there are $N$ query sets in the real query table with probability $p$, and $N^K-N$ sets with probability $pe^\epsilon$. Each $\epsilon$-deceptive query with respect to file index $k$ is chosen with equal probability to be sent to the databases as dummy queries at times $t_{i,j}$ when the file requirement at the corresponding time $T_i$ is $W_k$. Since there are $N-1$ deceptive queries,
\begin{align}
    P(Q_n=W_k^r|\theta=k,R=0)=\frac{1}{N-1}, \quad r\in\{1,\dotsc,N-1\}.
\end{align}
and
\begin{align}
    P(Q_n=W_k^r|\theta=j,R=0)=0, \quad r\in\{1,\dotsc,N-1\},\quad j\neq k.
\end{align}
for each database $n$, $n\in\{1,\dotsc,N\}$. Therefore, for all $\epsilon$-deceptive queries with respect to file index $k$ of the form $W_k^i$, the condition in \eqref{eq16} can be written as,
\begin{align}
    \frac{\alpha}{\alpha+\frac{1}{p(N-1)}(1-\alpha)}&=e^{-2\epsilon}
\end{align}
thus,
\begin{align}
    \alpha&=\frac{1}{p(N-1)(e^{2\epsilon}-1)+1}=\frac{N+(N^K-N)e^{\epsilon}}{(N-1)e^{2\epsilon}+(N^K-N)e^{\epsilon}+1},
\end{align}
which characterizes $\alpha=\mathbb{E}\left[\frac{1}{M+1}\right]$. The information available to  database $n$, $n\in\{1,\dotsc,N\}$, is the overall probability of receiving each query for each file requirement of the user $P(Q_n=\Tilde{Q}|\theta=k)$, $k\in\{1,\dotsc,K\}$, given by,
\begin{align}\label{overall}
    P(Q_n=\Tilde{Q}|\theta=k)&=P(Q_n=\Tilde{Q}|\theta=k,R=1)P(R=1|\theta=k)\nonumber\\
    &\qquad+P(Q_n=\Tilde{Q}|\theta=k,R=0)P(R=0|\theta=k).
\end{align}
For $\epsilon$-deceptive queries with respect to file index $k$, i.e., $W_k^j$, $j\in\{1,\dotsc,N-1\}$, the overall probability in \eqref{overall} from the perspective of database $n$, $n\in\{1,\dotsc,N\}$, is given by,
\begin{align}\label{dbkno1}
    P(Q_n=W_k^j|\theta=\ell)&=\begin{cases}
        \alpha p+\frac{1-\alpha}{N-1}=\frac{e^{2\epsilon}}{(N-1)(e^{2\epsilon}-1)+N+(N^K-N)e^\epsilon}, & \ell=k\\
        \alpha pe^\epsilon=\frac{e^\epsilon}{(N-1)(e^{2\epsilon}-1)+N+(N^K-N)e^\epsilon}, & \ell\neq k.
    \end{cases}
\end{align}
The probability of sending the null query $\phi$ to database $n$, $n\in\{1,\dotsc,N\}$, for each file-requirement $k$, $k\in\{1,\dotsc,K\}$, is,
\begin{align}\label{dbkno2}
    P(Q_n=\phi|\theta=k)=\alpha p=\frac{1}{(N-1)(e^{2\epsilon}-1)+N+(N^K-N)e^\epsilon}.
\end{align}
For the rest of the PIR queries denoted by $\hat{Q}$, i.e., queries of the form $\sum_{s=1}^\ell W_{i_s}^{j_s}$ for $\ell\in\{2,\dotsc,K\}$, the overall probability in \eqref{overall}, known by each database $n$, $n\in\{1,\dotsc,N\}$ for each file requirement $k$, $k\in\{1,\dotsc,K\}$ is given by,
\begin{align}\label{dbkno3}
     P(Q_n=\hat{Q}|\theta=k)=\alpha pe^\epsilon=\frac{e^\epsilon}{(N-1)(e^{2\epsilon}-1)+N+(N^K-N)e^\epsilon}.
\end{align}
Based on the query received at a given time $t$, each database $n$, $n\in\{1,\dotsc,N\}$, calculates the aposteriori probability of the user-required file index being $k$, $k\in\{1,\dotsc,K\}$, using,
\begin{align}\label{postfinal}
    P(\theta=k|Q_n=\Tilde{Q})&=\frac{P(Q_n=\Tilde{Q}|\theta=k)P(\theta=k)}{P(Q_n=\Tilde{Q})}.
\end{align}

Since we assume uniform priors, i.e., $P(\theta=k)=\frac{1}{K}$ for all $k\in\{1,\dotsc,K\}$, the posteriors are directly proportional to $P(Q_n=\Tilde{Q}|\theta=k)$ for each $\Tilde{Q}$. Therefore, the databases predict the user-required file index for each query received using \eqref{predict} and \eqref{dbkno1}-\eqref{dbkno3}. For example, when the query $W_1^1$ is received, it is clear that the maximum $P(\theta=k|Q_n=W_1^1)$ in \eqref{predict} is obtained for $k=1$ from \eqref{dbkno1} and \eqref{postfinal}. The prediction corresponding to any query received is given in Table~\ref{dbpredicts3} along with the corresponding probability of choosing the given prediction.\footnote{The superscript $j$ in the first column of Table~\ref{dbpredicts3} corresponds to any index in the set $\{1,\dotsc.N-1\}$.} 

Based on the information in Table~\ref{dbpredicts3}, the probability of error when a database $n$, $n\in\{1,\dotsc,N\}$, receives the query $W_k^\ell$ at some time $T_i$ is given by,
\begin{align}
    P(\hat{\theta}^{[T_i]}_{W_k^\ell}\neq\theta^{[T_i]})&=P(\theta^{[T_i]}\neq k|Q_n^{[T_i]}=W_k^\ell)\\
    &=\sum_{j=1,j\neq k}^K P(\theta^{[T_i]}=j|Q_n^{[T_i]}=W_k^\ell)\\
    &=\frac{\sum_{j=1,j\neq k}^K P(Q^{[T_i]}_n=W_k^\ell|\theta^{[T_i]}=j)P(\theta^{[T_i]}=j)}{P(Q_n^{[T_i]}=W_k^\ell)}\\
    &=\frac{\frac{1}{K}pe^\epsilon(K-1)}{P(Q^{[T_i]}_n=W_k^\ell)},\label{exp1}
\end{align}
where \eqref{exp1} follows from the fact that the user sends real queries based on the probabilities $P(Q_n=\Tilde{Q}|\theta=k,R=1)$ at time $T_i$.

\begin{table}[t]
\begin{center}
\begin{tabular}{ |c|c|c|c|c|c|}
\hline
query $\Tilde{Q}$ & $P(\hat{\theta}_{\Tilde{Q}}=1)$ & $P(\hat{\theta}_{\Tilde{Q}}=2)$ & $P(\hat{\theta}_{\Tilde{Q}}=3)$  & $\dotsc$ & $P(\hat{\theta}_{\Tilde{Q}}=K)$ \\
\hline
$W_1^j$ & $1$ & $0$ & $0$ & $\dotsc$ & $0$\\ 
\hline
$W_2^j$  & $0$ & $1$ & $0$ & $\dotsc$ & $0$\\
\hline
$W_3^j$  & $0$ & $0$ & $1$ & $\dotsc$ & $0$\\
\hline
$\vdots$  & $\vdots$ & $\vdots$ & $\vdots$ &  $\vdots$ & $\vdots$\\
\hline
$W_K^j$  & $0$ & $0$ & $0$ & $\dotsc$ & $1$\\
\hline
other queries  & $\frac{1}{K}$ & $\frac{1}{K}$ & $\frac{1}{K}$  & $\dotsc$ & $\frac{1}{K}$\\
\hline
\end{tabular}
\end{center}
\caption{Probabilities of each database predicting the user-required file.}
\label{dbpredicts3}
\end{table}
For all other queries $\Tilde{Q}$, the corresponding probability of error is given by,
\begin{align}
    P(\hat{\theta}^{[T_i]}_{\Tilde{Q}}\neq\theta^{[T_i]})&\!=\!P(\hat{\theta}^{[T_i]}\neq\theta^{[T_i]}|Q^{[T_i]}_n\!=\!\Tilde{Q})\\    
    &\!=\!\frac{\sum_{j=1}^K \sum_{m=1,m\neq j}^K P(\hat{\theta}^{[T_i]}\!=\!m,\theta^{[T_i]}\!=\!j,Q_n^{[T_i]}\!=\!\Tilde{Q})}{P(Q^{[T_i]}_n\!=\!\Tilde{Q})}\\
    &\!=\!\frac{\sum_{j=1}^K \sum_{m=1,m\neq j}^K P(\hat{\theta}^{[T_i]}\!=\!m|\theta^{[T_i]}\!=\!j,Q_n^{[T_i]}\!=\!\Tilde{Q})P(Q^{[T_i]}_n\!=\!\Tilde{Q}|\theta^{[T_i]}\!=\!j)P(\theta^{[T_i]}\!=\!j)}{P(Q^{[T_i]}_n\!=\!\Tilde{Q})}\\
    &\!=\!\frac{1}{P(Q^{[T_i]}_n\!=\!\Tilde{Q})}\begin{cases}
        \frac{(K-1)p}{K}, & \text{if } \Tilde{Q}=\phi\\
        \frac{(K-1)pe^\epsilon}{K}, & \text{if $\Tilde{Q}$ of the form $\sum_{s=1}^\ell W_{i_s}^{j_s}$, $\ell\in\{2,\dotsc,K\}$} 
    \end{cases}\label{exp2}
\end{align}
where \eqref{exp2} follows from the fact that $\hat{\theta}^{[T_i]}$ is conditionally independent of $\theta^{[T_i]}$ given $Q$ from \eqref{predict}. The probability of error of each database's prediction is given by,
\begin{align}
    P_e&=\sum_{\Tilde{Q}}P(Q_n^{[T_i]}=\Tilde{Q})P(\hat{\theta}^{[T_i]}\neq\theta^{[T_i]}|Q^{[T_i]}=\Tilde{Q})\\
    &=\sum_{k=1}^K\sum_{\ell=1}^{N-1}P(Q_n^{[T_i]}=W_k^\ell)\frac{\frac{1}{K}pe^\epsilon(K-1)}{P(Q_n^{[T_i]}=W_k^\ell)}+P(Q_n^{[T_i]}=\phi)\frac{\frac{1}{K}(K-1)p}{P(Q^{[T_i]}_n=\phi)}\nonumber\\
    &\qquad+(N^K-1-K(N-1))P(P(Q_n^{[T_i]}=\hat{Q})\frac{\frac{1}{K}(K-1)pe^\epsilon}{P(Q_n^{[T_i]}=\hat{Q})})\label{qhat}\\
    &=pe^\epsilon (K-1)(N-1)+\frac{(K-1)p}{K}+\frac{(K-1)pe^\epsilon(N^K-1-K(N-1))}{K}\\
    &=\frac{(K-1)(1+e^\epsilon(N^K-1))}{K(N+(N^K-N)e^\epsilon)},
\end{align}
where $\hat{Q}$ in \eqref{qhat} represents the queries of the form $\sum_{s=1}^\ell W_{i_s}^{j_s}$ for $\ell\in\{2,\dotsc,K\}$. Note that $P(Q_n^{[T_i]}=\hat{Q})$ is the same for each $\hat{Q}$ as $P(Q_n^{[T_i]}=\hat{Q}|\theta=j)=pe^\epsilon$ for each $\hat{Q}$ and all $j\in\{1,\dotsc,K\}$ from \eqref{pirquery}. Thus, the amount of deception achieved by this scheme for a given $\epsilon$ is given by,
\begin{align}
    D=P_e-\left(1-\frac{1}{K}\right)=\frac{(K-1)(N-1)(e^\epsilon-1)}{K(N+(N^K-N)e^\epsilon)}.
\end{align}
Therefore, for a required amount of deception $d$, satisfying $d<\frac{(K-1)(N-1)}{K(N^K-N)}$, the value of $\epsilon$ must be chosen as,
\begin{align}
    \epsilon=\ln\left(\frac{dKN+(K-1)(N-1)}{dKN+(K-1)(N-1)-dKN^K}\right).
\end{align}

The download cost of the general scheme is,
\begin{align}
    D_L&=\frac{1}{L}\left(NpL+(N^K-N)pe^\epsilon\frac{NL}{N-1}+\frac{NL}{N-1}\mathbb{E}[M]\right)\\
    D_L&=Np+\frac{N(N^K-N)}{N-1}pe^\epsilon+\left(\frac{N}{N-1}\right)\mathbb{E}[M]\\
    D_L&=\frac{N}{N-1}\left(1-\frac{1}{N+(N^K-N)e^\epsilon}+\mathbb{E}[M]\right).\label{dc}
\end{align}
Following optimization problem needs to be solved to minimize the download cost while satisfying $\alpha=\frac{N+(N^K-N)e^{\epsilon}}{(N-1)e^{2\epsilon}+(N^K-N)e^{\epsilon}+1}$, from \eqref{expect},
\begin{align}
    \min & \quad \mathbb{E}[M]\nonumber\\
    \text{s.t.} & \quad  \mathbb{E}\left[\frac{1}{M+1}\right]=\frac{N+(N^K-N)e^{\epsilon}}{(N-1)e^{2\epsilon}+(N^K-N)e^{\epsilon}+1}=\alpha.\label{opt}
\end{align}
\begin{lemma}\label{Lemma1}
The solution to the optimization problem in \eqref{opt} is given by,
\begin{align}\label{lemma}
    \mathbb{E}[M]=2u-u(u+1)\alpha,
\end{align}
where $u=\lfloor\frac{1}{\alpha}\rfloor$ for a given value of $\alpha$, which is specified by the required level of deception $d$.
\end{lemma}

The proof of Lemma~\ref{Lemma1} is given in the Appendix. The minimum download cost for the general case with $N$ databases, $K$ files and a deception requirement $d$, is obtained by \eqref{dc} and \eqref{lemma}. The corresponding maximum achievable rate is given in \eqref{main}.

\section{Discussion and Conclusions}

We introduced the problem of deceptive information retrieval (DIR), in which a user retrieves a file from a set of independent files stored in multiple databases, while revealing fake information about the required file to the databases, which makes the probability of error of the databases' prediction on the user-required file index high. The proposed scheme achieves rates lower than the PIR capacity when the required level of deception is positive, as it sends dummy queries at distinct time instances to deceive the databases. When the required level of deception is zero, the achievable DIR rate is the same as the PIR capacity.

The probability of error of the databases' prediction on the user-required file index is calculated at the time of the user's requirement, as defined in Section~\ref{formulate}. In the proposed scheme, the user sends dummy queries at other (future) time instances as well. As the databases are unaware of being deceived, and are unable to distinguish between the times corresponding to real and dummy queries, they make predictions on the user-required file indices every time a query is received. Note that whenever a query of the form $W_k^\ell$ is received, the databases prediction is going to be $\hat{\theta}=k$ from Table~\ref{dbpredicts3}. Although this is an incorrect prediction with high probability at times corresponding to user's real requirements, these predictions are correct when $W_k^\ell$ is used as a dummy query, as $W_k^\ell$ is only sent as a dummy query when the user requires to download file $k$. However, the databases are only able to obtain these correct predictions at future time instances, after which the user has already downloaded the required file while also deceiving the databases. 

The reason for the requirement of the time dimension is also explained as follows. An alternative approach to using the time dimension is to select a subset of databases to send the dummy queries and to send the real queries to rest of the databases. As explained above, whenever a database receives a query of the form $W_k^\ell$ as a dummy query, the database predicts the user-required file correctly. Therefore, this approach leaks information about the required file to a subset of databases, right at the time of the retrieval, while deceiving the rest. Hence, to deceive all databases at the time of retrieval, we exploit the time dimension that is naturally present in information retrieval applications that are time-sensitive.

A potential future direction of this work is an analysis on the time dimension. Note that in this work we assume that the databases do not keep track of the previous queries and only store the information corresponding to the current time instance. Therefore, as long as the dummy queries are sent at distinct time instances that are also different from the time of the user's requirement, the calculations presented in this paper are valid. An extension of basic DIR can be formulated by assuming that the databases keep track of all queries received and their time stamps. This imposes additional constraints on the problem as the databases now have extra information along the time dimension, which requires the scheme to choose the time instances at which the dummy queries are sent, in such a way that they do not leak any information about the existence of the two types (real and dummy) queries. Another direction is to incorporate the freshness and age of information into DIR, where the user may trade the age of the required file for a reduced download cost, by making use of the previous dummy downloads present in DIR. 

\appendix

\section{Proof of Lemma~\ref{Lemma1} }\label{app}

The solution to the optimization problem in \eqref{opt} for the general case with $N$ databases and $K$ files is as follows. The optimization problem in \eqref{opt}, for a required amount of deception $d$ and the corresponding $\epsilon$ with $\alpha=\frac{N+(N^K-N)e^{\epsilon}}{(N-1)e^{2\epsilon}+(N^K-N)e^{\epsilon}+1}$ is given by,
\begin{align}
     \min & \quad\mathbb{E}[M]=\sum_{m=0}^{\infty} mp_m\nonumber\\
    \text{s.t.} & \quad \mathbb{E}\left[\frac{1}{m+1}\right]=\sum_{m=0}^{\infty} \left(\frac{1}{m+1}\right)p_m = \alpha \nonumber\\
    &\qquad \sum_{m=0}^{\infty} p_m=1\nonumber\\
    &\qquad \ p_m \geq 0, \quad m\in\{0,1,\dotsc\}.
\end{align}

We need to determine the optimum PMF of $M$ that minimizes $\mathbb{E}[M]$ while satisfying the given condition. The Lagrangian $L$ of this optimization problem is given by,
\begin{align}
     L=\sum_{m=0}^{\infty} mp_m+\lambda_1\left(\sum_{m=0}^{\infty} \left(\frac{1}{m+1}\right)p_m-\alpha\right)+\lambda_2\left(\sum_{m=0}^{\infty} p_m-1\right)-\sum_{m=0}^{\infty}\mu_mp_m.
\end{align}
Then, the following set of equations need to be solved to find the minimum $\mathbb{E}[M]$,
\begin{align}\label{kkt1}
     \frac{\partial L}{\partial p_m}=m+\lambda_1\left(\frac{1}{m+1}\right)+\lambda_2-\mu_m&=0,\quad m\in\{0,1,\dotsc\}\\
    \sum_{m=0}^{\infty} \left(\frac{1}{m+1}\right)p_m&=\alpha\label{eq66}\\
    \sum_{m=0}^{\infty} p_m&=1\label{sum1}\\
    \mu_mp_m&=0, \quad m\in\{0,1,\dotsc\}\\
    \mu_m,p_m &\geq 0, \quad m\in\{0,1,\dotsc\}.
    \label{kkt2}
\end{align}

\textbf{Case~1:} Assume that the PMF of $M$ contains at most two non-zero probabilities, i.e., $p_0,p_1\geq0$ and $p_i=0$, $i\in\{2,3,\dotsc\}$. Then, the conditions in \eqref{kkt1}-\eqref{kkt2} are simplified as, 
\begin{align}
    \frac{\partial L}{\partial p_0}=\lambda_1+\lambda_2-\mu_0&=0\\
    \frac{\partial L}{\partial p_1}=\frac{1}{2}\lambda_1+\lambda_2-\mu_1&=-1\\
    p_0+\frac{1}{2}p_1&=\alpha\label{eq72}\\
    p_0+p_1&=1\label{eq73}\\
    \mu_0p_0&=0\\
    \mu_1p_1&=0\\
    \mu_0,\mu_1,p_0,p_1 &\geq 0\label{eq76}.
\end{align}
From \eqref{eq72} and \eqref{eq73} we obtain,
\begin{align}
    p_0+\frac{1}{2}(1-p_0)&= \alpha
\end{align}
and thus,
\begin{align}
    p_0= 2\alpha-1, \qquad p_1=2-2\alpha,
\end{align}
which along with \eqref{eq76} implies that this solution is only valid for $\frac{1}{2}\leq \alpha\leq 1$. The corresponding optimum value of $\mathbb{E}[M]$ is given by,
\begin{align}
    \mathbb{E}[M]=1-p_0=2-2\alpha, \qquad \frac{1}{2}\leq \alpha\leq1.
\end{align}

\textbf{Case~2:} Now consider the case where at most three probabilities of the PMF of $M$ are allowed to be non-zero. i.e., $p_0,p_1,p_2\geq0$ and $p_i=0$, $i\in\{3,4,\dotsc\}$. The set of conditions in \eqref{kkt1}-\eqref{kkt2} for this case is,
\begin{align}
\label{c21}
    \frac{\partial L}{\partial p_m}=m+\lambda_1\left(\frac{1}{m+1}\right)+\lambda_2-\mu_m&=0,\quad m\in\{0,1,2\}\\
    \sum_{m=0}^{2} \left(\frac{1}{m+1}\right)p_m&=\alpha\label{eq82}\\
    \sum_{m=0}^{2} p_m&=1\label{eq83}\\
    \mu_mp_m&=0, \quad m\in\{0,1,2\}\label{eq84}\\
    \mu_m,p_m &\geq 0, \quad m\in\{0,1,2\}\label{c22}.
\end{align}
The set of conditions in \eqref{c21}-\eqref{c22} can be written in a matrix form as,
\begin{align}
\begin{bmatrix}
    1 & 1 & -1 & 0 & 0 & 0 & 0 & 0\\
    \frac{1}{2} & 1 & 0 & -1 & 0 & 0 & 0 & 0\\
    \frac{1}{3} & 1 & 0 & 0 & -1 & 0 & 0 & 0\\
    0 & 0 & 0 & 0 & 0 & 1 & \frac{1}{2} & \frac{1}{3}\\
    0 & 0 & 0 & 0 & 0 & 1 & 1 & 1\\
\end{bmatrix}
\begin{bmatrix}
     \lambda_1\\\lambda_2\\\mu_0\\ \mu_1\\\mu_2\\p_0\\p_1\\p_2  
\end{bmatrix}
=
\begin{bmatrix}
0\\-1\\-2\\\alpha\\1\label{eq86}
\end{bmatrix}.
\end{align}
Three of the above eight variables, i.e., either $\mu_i$ or $p_i$ for each $i$, are always zero according to \eqref{eq84}. We consider all choices of $\{\mu_i,p_i\}$ pairs such that one element of the pair is equal to zero, and the other one is a positive variable, and solve the system for the non-zero variables. Then we calculate the resulting  $\mathbb{E}[M]$, along with the corresponding regions of $u$ for which the solutions are applicable. For each region of $u$, we find the solution to \eqref{eq86} that results in the minimum $\mathbb{E}[M]$. Based on this process, the optimum values of $p_i$, $i\in\{0,1,2\}$, the corresponding ranges of $u$ and the minimum values of $\mathbb{E}[M]$ are given in Table~\ref{case2res}. 

\begin{table}[h]
\begin{center}
\begin{tabular}{ |c|c|c|c|c|}
\hline
range of $\alpha$ & $p_0$ & $p_1$ & $p_2$  & $\mathbb{E}[M]$ \\
\hline
$\frac{1}{3}\leq \alpha\leq\frac{1}{2}$ & $0$ & $6\alpha-2$ & $3-6\alpha$ & $4-6\alpha$\\ 
\hline
$\frac{1}{2}\leq \alpha\leq1$ & $2\alpha-1$ & $2-2\alpha$ & $0$ & $2-2\alpha$\\
\hline
\end{tabular}
\end{center}
\caption{Solution to Case~2: Optimum PMF of $M$, valid ranges of $\alpha$ and minimum $\mathbb{E}[M]$.}
\label{case2res}
\end{table}

As an example, consider the calculations corresponding to the case where $\mu_0>0$, $\mu_1=\mu_2=0$ which implies $p_0=0$, $p_1,p_2>0$. Note that for this case, \eqref{eq86} simplifies to,
\begin{align}
\begin{bmatrix}
    1 & 1 & -1 & 0 & 0\\
    \frac{1}{2} & 1 & 0 & 0 & 0\\
    \frac{1}{3} & 1 & 0 & 0 & 0\\
    0 & 0 & 0 & \frac{1}{2} & \frac{1}{3}\\
    0 & 0 & 0 & 1 & 1\\
\end{bmatrix}
\begin{bmatrix}
     \lambda_1\\\lambda_2\\\mu_0\\p_1\\p_2  
\end{bmatrix}
=
\begin{bmatrix}
0\\-1\\-2\\\alpha\\1\label{eq87}
\end{bmatrix}.
\end{align}
The values of $p_1$ and $p_2$, from the solution of the above system, and the corresponding range of $\alpha$, from \eqref{c22}, along with the resulting $\mathbb{E}[M]$ are given by,
\begin{align}
  p_1=6\alpha-2, \quad p_2=3-6\alpha, \quad \frac{1}{3}\leq \alpha \leq \frac{1}{2}, \quad \mathbb{E}[M]=4-6\alpha.   
\end{align}

\textbf{Case~3:} At most four non-zero elements of the PMF of $M$ are considered in this case, i.e., $p_0,p_1,p_2,p_3\geq0$ and $p_i=0$, $i\in\{4,5,\dotsc\}$. The conditions in \eqref{kkt1}-\eqref{kkt2} can be written in a matrix form as,
\begin{align}
\begin{bmatrix}
    1 & 1 & -1 & 0 & 0 & 0 & 0 & 0 & 0 & 0\\
    \frac{1}{2} & 1 & 0 & -1 & 0 & 0 & 0 & 0 & 0 & 0\\
    \frac{1}{3} & 1 & 0 & 0 & -1 & 0 & 0 & 0 & 0 & 0\\
    \frac{1}{4} & 1 & 0 & 0 & 0 & -1 & 0 & 0 & 0 & 0 \\
    0 & 0 & 0 & 0 & 0 & 0 & 1 & \frac{1}{2} & \frac{1}{3} & \frac{1}{4}\\
    0 & 0 & 0 & 0 & 0 & 0 & 1 & 1 & 1 & 1\\
\end{bmatrix}
\begin{bmatrix}
     \lambda_1\\\lambda_2\\\mu_0\\ \mu_1\\\mu_2\\\mu_3\\p_0\\p_1\\p_2\\p_3  
\end{bmatrix}
=
\begin{bmatrix}
0\\-1\\-2\\-3\\\alpha\\1\label{eq89}
\end{bmatrix}.
\end{align}
Using the same method described in Case~2, the optimum values of $p_i$, $i\in\{0,1,2,3\}$, corresponding ranges of $\alpha$ and the resulting minimum $\mathbb{E}[M]$ for Case~3 are given in Table~\ref{case3res}.

\begin{table}[h]
\begin{center}
\begin{tabular}{ |c|c|c|c|c|c|}
\hline
range of $\alpha$ & $p_0$ & $p_1$ & $p_2$ & $p_3$  & $\mathbb{E}[M]$ \\
\hline
$\frac{1}{4}\leq \alpha\leq\frac{1}{3}$ & $0$ & $0$ & $12\alpha-3$ & $4-12\alpha$ & $6-12\alpha$\\ 
\hline
$\frac{1}{3}\leq \alpha\leq\frac{1}{2}$ & $0$ & $6\alpha-2$ & $3-6\alpha$ & $0$ & $4-6\alpha$\\ 
\hline
$\frac{1}{2}\leq \alpha\leq1$ & $2\alpha-1$ & $2-2\alpha$ & $0$ & $0$ & $2-2\alpha$\\ 
\hline
\end{tabular}
\end{center}
\caption{Solution to Case~3: Optimum PMF of $M$, valid ranges of $\alpha$ and minimum $\mathbb{E}[M]$.}
\label{case3res}
\end{table}

\textbf{Case~4:} At most five non-zero elements of the PMF of $M$ are considered in this case, i.e., $p_0,p_1,p_2,p_3,p_4\geq0$ and $p_i=0$, $i\in\{5,6,\dotsc\}$. The conditions in \eqref{kkt1}-\eqref{kkt2} can be written in a matrix form as,
\begin{align}
\begin{bmatrix}
    1 & 1 & -1 & 0 & 0 & 0 & 0 & 0 & \dotsc & 0\\
    \frac{1}{2} & 1 & 0 & -1 & 0 & 0 & 0 & 0 & \dotsc & 0\\
    \frac{1}{3} & 1 & 0 & 0 & -1 & 0 & 0 & 0 & \dotsc & 0\\
    \frac{1}{4} & 1 & 0 & 0 & 0 & -1 & 0 & 0 & \dotsc & 0 \\
    \frac{1}{5} & 1 & 0 & 0 & 0 & 0 & -1 & 0 & \dotsc & 0 \\
    0 & \dotsc & 0 & 0 & 0 & 1 & \frac{1}{2} & \frac{1}{3} & \frac{1}{4} & \frac{1}{5}\\
    0 & \dotsc & 0 & 0 & 0 & 1 & 1 & 1 & 1 & 1\\
\end{bmatrix}
\begin{bmatrix}
     \lambda_1\\\lambda_2\\\mu_0\\ \mu_1\\\mu_2\\\mu_3\\\mu_4\\p_0\\p_1\\p_2\\p_3\\p_4  
\end{bmatrix}
=
\begin{bmatrix}
0\\-1\\-2\\-3\\-4\\\alpha\\1\label{eq88}
\end{bmatrix}.
\end{align}

Using the same method as before, the optimum values of $p_i$, $i\in\{0,1,2,3,4\}$, the corresponding ranges of $\alpha$ and the resulting minimum $\mathbb{E}[M]$ for Case~4 are given in Table~\ref{case4res}.

\begin{table}[h]
\begin{center}
\begin{tabular}{ |c|c|c|c|c|c|c|}
\hline
range of $\alpha$ & $p_0$ & $p_1$ & $p_2$ & $p_3$ & $p_4$  & $\mathbb{E}[M]$ \\
\hline
$\frac{1}{5}\leq \alpha\leq\frac{1}{4}$ & $0$ & $0$ & $0$ & $20\alpha-4$ & $5-20\alpha$ & $8-20\alpha$\\ 
\hline
$\frac{1}{4}\leq \alpha\leq\frac{1}{3}$ & $0$ & $0$ & $12\alpha-3$ & $4-12\alpha$ & $0$ & $6-12\alpha$\\ 
\hline
$\frac{1}{3}\leq \alpha\leq\frac{1}{2}$ & $0$ & $6\alpha-2$ & $3-6\alpha$ & $0$ & $0$ & $4-6\alpha$\\
\hline
$\frac{1}{2}\leq \alpha\leq1$ & $2\alpha-1$ & $2-2\alpha$ & $0$ & $0$ & $0$ & $2-2\alpha$\\
\hline
\end{tabular}
\end{center}
\caption{Solution to Case~4: Optimum PMF of $M$, valid ranges of $\alpha$ and minimum $\mathbb{E}[M]$.}
\label{case4res}
\end{table}

Note that the PMF of $M$ and the resulting $\mathbb{E}[M]$ are the same for a given $\alpha$ in all cases (see Tables~\ref{case2res}-\ref{case4res}) irrespective of the support of the PMF of $M$ considered. Therefore, we observe from the above cases that, for a given $\alpha$ in the range $\frac{1}{\ell+1}\leq \alpha\leq\frac{1}{\ell}$, $\mathbb{E}[M]$ is minimized when the PMF of $M$ is such that,
\begin{align}
    p_\ell,p_{\ell-1}>0, \text{ and } p_i=0 \text{ for } i\in\mathbb{Z^+}\setminus\{\ell,\ell-1\},
\end{align}
which requires $p_{\ell}$ and $p_{\ell-1}$ to satisfy,
\begin{align}
    p_{\ell}+p_{\ell-1}&=1\\
    \mathbb{E}\left[\frac{1}{M+1}\right]=p_\ell\frac{1}{\ell+1}+p_{\ell-1}\frac{1}{\ell}&=\alpha.
\end{align}
Therefore, for a given $\alpha$ in the range $\frac{1}{\ell+1}\leq \alpha\leq\frac{1}{\ell}$, the optimum PMF of $M$ and the resulting minimum $\mathbb{E}[M]$ are given by,
\begin{align}
    p_\ell=(\ell+1)(1-\ell \alpha),\quad p_{\ell-1}=\ell((\ell+1)\alpha-1), \quad \mathbb{E}[M]=2\ell-\alpha\ell(\ell+1).
\end{align}

\bibliographystyle{unsrt}

\bibliography{references}

\begin{thebibliography}{10}

\bibitem{original}
B.~Chor, E.~Kushilevitz, O.~Goldreich, and M.~Sudan.
\newblock Private information retrieval.
\newblock {\em Journal of the ACM}, 45(6):965--981, November 1998.

\bibitem{PIR}
H.~Sun and S.~A. Jafar.
\newblock The capacity of private information retrieval.
\newblock {\em IEEE Transactions on Information Theory}, 63(7):4075--4088, July
  2017.

\bibitem{ChaoTian}
C.~Tian, H.~Sun, and J.~Chen.
\newblock Capacity-achieving private information retrieval codes with optimal
  message size and upload cost.
\newblock {\em IEEE Transactions on Information Theory}, 65(11):7613--7627,
  November 2019.

\bibitem{coded}
K.~Banawan and S.~Ulukus.
\newblock The capacity of private information retrieval from coded databases.
\newblock {\em IEEE Transactions on Information Theory}, 64(3):1945--1956,
  March 2018.

\bibitem{colluding}
H.~Sun and S.~A. Jafar.
\newblock The capacity of robust private information retrieval with colluding
  databases.
\newblock {\em IEEE Transactions on Information Theory}, 64(4):2361--2370,
  April 2018.

\bibitem{sideinfo}
S.~Kadhe, B.~Garcia, A.~Heidarzadeh, S.~El Rouayheb, and A.~Sprintson.
\newblock Private information retrieval with side information.
\newblock {\em IEEE Transactions on Information Theory}, 66(4):2032--2043,
  April 2020.

\bibitem{singleDB}
S.~Li and M.~Gastpar.
\newblock Single-server multi-message private information retrieval with side
  information: the general cases.
\newblock In {\em IEEE ISIT}, June 2020.

\bibitem{byzantine}
K.~Banawan and S.~Ulukus.
\newblock The capacity of private information retrieval from {B}yzantine and
  colluding databases.
\newblock {\em IEEE Transactions on Information Theory}, 65(2):1206--1219,
  February 2019.

\bibitem{SecureStorage}
H.~Yang, W.~Shin, and J.~Lee.
\newblock Private information retrieval for secure distributed storage systems.
\newblock {\em IEEE Transactions on Information Forensics and Security},
  13(12):2953--2964, December 2018.

\bibitem{XSTPIR}
Z.~Jia and S.~A. Jafar.
\newblock {$X$}-secure {$T$}-private information retrieval from {MDS} coded
  storage with {B}yzantine and unresponsive servers.
\newblock {\em IEEE Transactions on Information Theory}, 66(12):7427--7438,
  December 2020.

\bibitem{MMPIR}
K.~Banawan and S.~Ulukus.
\newblock Multi-message private information retrieval: Capacity results and
  near-optimal schemes.
\newblock {\em IEEE Transactions on Information Theory}, 64(10):6842--6862,
  October 2018.

\bibitem{evesdroppers}
Q.~Wang, H.~Sun, and M.~Skoglund.
\newblock The capacity of private information retrieval with eavesdroppers.
\newblock {\em IEEE Transactions on Information Theory}, 65(5):3198--3214, May
  2019.

\bibitem{leaky}
I.~Samy, M.~Attia, R.~Tandon, and L.~Lazos.
\newblock Asymmetric leaky private information retrieval.
\newblock {\em IEEE Transactions on Information Theory}, 67(8):5352--5369,
  August 2021.

\bibitem{ChaoTian_leakage}
T.~Guo, R.~Zhou, and C.~Tian.
\newblock On the information leakage in private information retrieval systems.
\newblock {\em IEEE Transactions on Information Forensics and Security},
  15:2999--3012, December 2020.

\bibitem{cyber_defense}
D.~Liebowitz, S.~Nepal, K.~Moore, C.~Christopher, S.~Kanhere, D.~Nguyen,
  R.~Timmer, M.~Longland, and K.~Rathakumar.
\newblock Deception for cyber defence: Challenges and opportunities.
\newblock In {\em TPS-ISA}, December 2021.

\bibitem{cybersecurity}
A.~Yarali and F.~Sahawneh.
\newblock Deception: Technologies and strategy for cybersecurity.
\newblock In {\em SmartCloud}, December 2019.

\bibitem{adaptivedeception}
C.~Faveri and A.~Moreira.
\newblock Designing adaptive deception strategies.
\newblock In {\em QRS-C}, August 2016.

\bibitem{proofofconcept}
W.~Tounsi.
\newblock Cyber deception, the ultimate piece of a defensive strategy - proof
  of concept.
\newblock In {\em CSNet}, October 2022.

\bibitem{diversity}
A.~Sarr, A.~Anwar, C.~Kamhoua, N.~Leslie, and J.~Acosta.
\newblock Software diversity for cyber deception.
\newblock In {\em IEEE Globecom}, December 2020.

\bibitem{ravi-leaky}
I.~{Samy}, R.~{Tandon}, and L.~{Lazos}.
\newblock On the capacity of leaky private information retrieval.
\newblock In {\em IEEE ISIT}, July 2019.

\bibitem{semanticPIR}
S.~Vithana, K.~Banawan, and S.~Ulukus.
\newblock Semantic private information retrieval.
\newblock {\em IEEE Transactions on Information Theory}, 68(4):2635--2652,
  April 2022.

\end{thebibliography}

\end{document}